\documentclass[11pt]{article}
\font\af=msbm10

\newcommand{\be}{\begin{equation}}
\newcommand{\ee}{\end{equation}}
\newcommand{\bea}{\begin{eqnarray}}
\newcommand{\eea}{\end{eqnarray}}
\newcommand{\ben}{\begin{eqnarray*}}
\newcommand{\een}{\end{eqnarray*}}
\newcommand{\ds}{\displaystyle}

\newcommand{\myinta}{\int\!\!d^3\!x\,} 
\newcommand{\myintb}{\int\!\!d^3\!x'\,}
\newcommand{\pa}{\partial}
\newcommand{\na}{\nabla}
\newcommand{\ve}{\varepsilon}

\def\pp{{\bf p}^2}
\def\ppp{({\bf p}^2)}
\def\dpdp{{\bf q}^2}
\def\pdp{({\bf p}\cdot{\bf q})}
\def\drdr{{\bf v}^2}
\def\dpdr{({\bf q}\cdot{\bf v})}
\def\pdr{({\bf p}\cdot{\bf v})}
\def\np{({\bf n}\cdot{\bf p})}
\def\ndp{({\bf n}\cdot{\bf q})}
\def\ndr{({\bf n}\cdot{\bf v})}

\def\pasq{{\bf p}_a^2}
\def\pbsq{{\bf p}_b^2}
\def\pasqp{\left({\bf p}_a^2\right)}
\def\papb{\left({\bf p}_a\cdot{\bf p}_b\right)}
\def\nanb{\left(\na_a\cdot\na_b\right)}
\def\pana{\left({\bf p}_a\cdot\na_a\right)}
\def\pbnb{\left({\bf p}_b\cdot\na_b\right)}
\def\panb{\left({\bf p}_a\cdot\na_b\right)}
\def\pbna{\left({\bf p}_b\cdot\na_a\right)}
\def\nna{\left(\na\cdot\na_a\right)}
\def\nnb{\left(\na\cdot\na_b\right)}
\def\pan{\left({\bf p}_a\cdot\na\right)}
\def\pbn{\left({\bf p}_b\cdot\na\right)}

\def\uapa{\left({\bf n}_a\cdot{\bf p}_a\right)}
\def\uabpa{\left({\bf n}_{ab}\cdot{\bf p}_a\right)}

\def\piti#1#2{\widetilde{\pi}^{#1#2}}
\def\pitiii#1#2{\widetilde{\pi}^{#1#2}_{(3)}}
\def\pitv#1#2{\widetilde{\pi}^{#1#2}_{(5)}}
\def\pitvii#1#2{\widetilde{\pi}^{#1#2}_{(7)}}
\def\gpitiii#1#2#3{\widetilde{\pi}^{#1#2}_{(3),#3}}
\def\gpitv#1#2#3{\widetilde{\pi}^{#1#2}_{(5),#3}}
\def\gpitvii#1#2#3{\widetilde{\pi}^{#1#2}_{(7),#3}}

\def\pitt#1#2{\pi^{#1#2{\rm TT}}}
\def\pittv#1#2{\pi^{#1#2{\rm TT}}_{(5)}}

\def\htt#1#2{h^{\rm TT}_{#1#2}}
\def\httiv#1#2{h^{\rm TT}_{(4)#1#2}}
\def\httv#1#2{h^{\rm TT}_{(5)#1#2}}
\def\httvi#1#2{h^{\rm TT}_{(6)#1#2}}

\def\httivdot#1#2{\dot{h}^{\rm TT}_{(4)#1#2}}

\def\ghtt#1#2#3{h^{\rm TT}_{#1#2,#3}}
\def\ghttiv#1#2#3{h^{\rm TT}_{(4)#1#2,#3}}
\def\ghttvi#1#2#3{h^{\rm TT}_{(6)#1#2,#3}}

\begin{document}

\title{
3rd post-Newtonian higher order ADM Hamilton dynamics\\
for two-body point-mass systems}

\author{Piotr Jaranowski
\\{\it Institute of Physics, Bia{\l}ystok University} 
\\{\it Lipowa 41, 15-424 Bia{\l}ystok, Poland}
\thanks{E-mail: pio@alpha.fuwb.edu.pl}
\and Gerhard Sch\"afer
\\{\it Theoretisch-Physikalisches Institut,
Friedrich-Schiller-Universit\"at}
\\{\it Max-Wien-Platz 1, 07743 Jena, Germany}
\thanks{E-mail: gos@tpi.uni-jena.de}}

\date{}

\maketitle

\begin{abstract}

The paper presents the conservative dynamics of two-body point-mass systems up
to the third post-Newtonian order ($1/c^6$). The two-body dynamics is given in
terms of a higher order ADM Hamilton function which results from a third
post-Newtonian Routh functional for the total field-plus-matter system. The
applied regularization procedures, together with making use of distributional
differentiation of homogeneous functions, give unique results for the terms in
the Hamilton function apart from the coefficient of the term $(\nu
p_{i}{\pa_{i}})^2r^{-1}$. The result suggests an invalidation of the binary
point-mass model at the third post-Newtonian order.

\vspace{0.5cm}\noindent PACS number(s): 04.25.Nx, 04.30.Db, 97.60.Jd, 97.60.Lf
\end{abstract}

\section{Introduction and summary}

The calculation of general relativistic equations of motion for compact binary
systems, in the recent past, was exclusively devoted to the obtention of higher
order post-Newtonian gravitational radiation reaction contributions. With the
works by Iyer and Will \cite{IW95}, Blanchet \cite{B97}, and the authors
\cite{JS97} the radiation reaction levels have been completed up to 3.5
post-Newtonian (3.5PN) order, i.e.\ to the order $(1/c^2)^{7/2}$ beyond the
Newtonian dynamics (for the 2.5PN order see, e.g., the review \cite{D87a}).
Quite recently, Gopakumar et al.\ \cite{GII97} succeeded in giving first
results for the gravitational radiation reaction in compact binary systems at
4.5PN order, applying balance equations between far-zone fluxes and near-zone
losses of energy and angular momentum. The conservative 3PN and 4PN orders of
approximation, however, have not been tackled so far (up to the 2PN level of
approximation see, e.g., the review \cite{D87} and the paper \cite{DS88}).

It is the aim of the present paper to develop the two-body point-mass dynamics
up to the 3PN order of approximation using the canonical formalism of
Arnowitt-Deser-Misner (ADM) \cite{ADM62}. There are several aspects which make
such a calculation rather interesting: (i) the well-known problem of
applicability of Dirac delta distributions for the source of the general
relativistic gravitational field; (ii) the need of the 3PN dynamics for a
better undertanding of the innermost stable circular orbit for binary systems
\cite{KWW92/93,WS93,SW93}; (iii) the importance of the 3PN dynamics to control
the viability of the 2PN filters for gavitational wave measurements from
inspiralling compact binaries \cite{DIS97}.

Within our canonical approach, together with applying Hadamard's ``partie finie'' technic,
some specific analytic regularization formula and a generalization of it achieved
by us --- the non-generalized formula has been successfully used by
Damour \cite{D83}, Damour and Sch\"afer \cite{DS85}, and Kopeikin \cite{K85} at
the 2PN level and by us \cite{JS97} at the 3.5PN level ---, and the
distributional differentiation of homogeneous functions, we were able to
explicitly calculate and uniquely regularize all terms which occur at the 3PN
order of approximation, but one (notice: this is the only ambiguous term up to
the 3.5PN order of approximation, inclusively). The coefficient of the
following term turns out to be finite but ambiguous: $-(\nu p_i \pa_i)^2r^{-1}$,
or, in non-reduced-variable form (apart from a factor $4$):
$$
\frac{m_1v^i_1v^j_1}{2c^2} r^2_{s1} \pa_{1i}\pa_{1j}
\left(-\frac{G m_2}{r_{12}}\right) +
\frac{m_2v^i_2v^j_2}{2c^2} r^2_{s2} \pa_{2i}\pa_{2j}
\left(-\frac{G m_1}{r_{12}}\right),
$$
where $m_1, v^i_1, r_{s1}$ and $m_2, v^i_2, r_{s2}$ denote the masses,
velocities, and Schwarzschild radii ($r_{sa} = Gm_a/2c^2$, in isotropic
coordinates) of the bodies 1 and 2, respectively,
and $r_{12}$ their relative distance. This term describes the
quadrupole-quadrupole interaction of the kinetic energy tensor of each body
with the gravitational tidal field of the other body scaled to the
Schwarzschild radius of the former one. The bodies, in this interaction term,
are obviously not tidally deformed (this is expected from Ref.\ \cite{D87}
where it is shown that the tidal deformation comes in at 5PN only) but they
seem to have been attributed extensions of Schwarzschild radius size. Those
extensions are beyond the applicability of the 3PN approximation so the
obtained result may suggest that a fully consistent 3PN compact binary model
needs extended bodies as source for the gravitational field. Only within the
complete theory the bodies' Schwarzschild radii are expected to
enter the equations for the binary system in a consistent and unique manner.

The paper is  organized as follows. In Section 2 we introduce the point-mass
model and the post-Newtonian approximation scheme and we develop the constraint
equations to the 3PN order of approximation. To obtain an autonomous Hamilton
function for the bodies at 3PN order, dropping the dissipative 2.5PN level, we
introduce in Section 3 the Routh functional for the total field-plus-matter
system and eliminate the field degrees of freedom for the bodies' variables.
The autonomous Hamilton function comes out of higher order. The Section 4 is
devoted to the explicit calculation of the higher order autonomous Hamilton
function applying the regularization procedures of the Appendices B.1 and B.2.
In the Section 5 a thourough investigation of the obtained ambiguity is
undertaken. In the Section 6 we compare our results with limiting expressions
known from the literature. The Appendix A presents several explicit metric
coefficients and some useful formulae for inverse Laplacians. The Appendix B.1
is devoted to the Hadamard's ``partie finie'' regularization. The Appendix B.2
presents a powerful analytic regularization procedure based on an analytic
formula derived by Riesz and generalized by us in this paper. In the Appendix
B.3 the analytic regularization procedure devised by Riesz is given. The
Appendix B.4 shows the distributional differentiation of homogeneous functions.

We use units in which $16\pi G=c=1$, where $G$ is the Newtonian gravitational
constant and $c$ the velocity of light.  We employ the following notation:
${\bf x}=\left(x^i\right)$ ($i=1,2,3$) denotes a point in the 3-dimensional
Euclidean space {\af R}$^3$ endowed with a standard Euclidean metric and a
scalar product (denoted by a dot).  Letters $a$ and $b$ are body labels,
so ${\bf x}_a\in\mbox{\af R}^3$ denotes the position of the $a$th point mass.
We also define  ${\bf r}_a := {\bf x} - {\bf x}_a$, $r_a := |{\bf r}_a|$,
${\bf n}_a := {\bf r}_a/r_a$; and for $a\ne b$,  ${\bf r}_{ab} := {\bf x}_a -
{\bf x}_b$, $r_{ab} := |{\bf r}_{ab}|$, ${\bf n}_{ab} := {\bf r}_{ab}/r_{ab}$;
$|\cdot|$ stands here for the length of a vector.  The linear momentum vector of the
$a$th body is denoted by ${\bf p}_a=\left(p_{ai}\right)$, and $m_a$
denotes its mass parameter.  Indices with round brackets, like in
$\phi_{(2)}$, give the order of the object in inverse powers of the velocity
of light, in this case, $1/c^2$.  We abbreviate $\delta\left({\bf x}-{\bf
x}_a\right)$ by $\delta_a$. An overdot, like in $\dot{\bf x}_a$, means the
total time derivative. The partial differentiation with respect to $x^i$ is
denoted by $\pa_i$ or by a comma, i.e., $\pa_i\phi\equiv\phi_{,i}$; the
partial differentiation with respect to $x_a^i$ we denote by $\pa_{ai}$.

Throughout this paper we extensively used the computer algebra system {\em
Mathematica} \cite{W91}.

\section{The constraint equations up to 3PN order}

We consider a many-body point-mass system which interacts with the
gravitational field according to the theory of general relativity. For such a
system the constraint equations in the canonical formalism of ADM read (see,
e.g., Eqs.\ (2.8) in \cite{S85})
\bea
\label{ce1}
g^{-1/2}\left[
gR+\frac{1}{2}\left(g_{ij}\pi^{ij}\right)^2-\pi_{ij}\pi^{ij}\right]
&=&\sum_a\left(g^{ij}p_{ai}p_{aj}+m_a^2\right)^{1/2}\delta_a,
\\[2ex]
\label{ce2}
-2{\pi^{ij}}_{|j}&=&\sum_a g^{ij}p_{aj}\delta_a.
\eea
Here $g_{ij}={^4g}_{ij}$ are the field variables (the prefix ``$^4$" denotes
a four-dimensional quantity, all unmarked quantities are understood as
three-dimensional), $g^{ij}$ is the inverse of $g_{ij}$
($g^{ij}g_{jk}=\delta^i_k$), $g:=\det\{g_{ij}\}$, ``$ _|$" indicates the
covariant derivative with respect to $g_{ij}$, $R$ is the curvature scalar
formed from the metric $g_{ij}$; $\pi^{ij}$ is the canonical conjugate to the
field $g_{ij}$.  Spatial indices are raised and lowered using $g^{ij}$ and
$g_{ij}$, respectively.

We use the following coordinate conditions (see, e.g., Eqs.\ (2.9) in
\cite{S85})
\bea
\label{cc1}
g_{ij}&=&\left(1+\frac{1}{8}\phi\right)^4\delta_{ij}+{\htt ij},
\\[2ex]
\label{cc2}
\pi^{ii}&=&0,
\eea
where ${\htt ij}$ is the transverse traceless part of $g_{ij}-\delta_{ij}$.
The trace-free field momentum $\pi^{ij}$ can be split into two parts, a
longitudinal ${\piti ij}$ and a transversal ${\pitt ij}$ one:
\be
\pi^{ij}={\piti ij}+{\pitt ij},
\ee
where ${\piti ij}$ can be expressed in terms of a single vector $\pi^i$
as follows:
\be
\label{pitidec}
{\piti ij}=\pi^i_{,j}+\pi^j_{,i}-\delta_{ij}\pi^k_{,k}+\Delta^{-1}\pi^k_{,ijk}.
\ee

If both the constraint equations (\ref{ce1})--(\ref{ce2}) and the coordinate
conditions (\ref{cc1})--(\ref{cc2}) are satisfied, the Hamiltonian of the
theory can be put into its reduced form, which can be written as
\be
\label{hred}
H\left[{\bf x}_a,{\bf p}_{a},{\htt ij},{\pitt ij}\right]
=-\myinta\Delta\phi\left[{\bf x}_a,{\bf p}_{a},{\htt ij},{\pitt ij}\right].
\ee
The reduced Hamiltonian contains the full information for the dynamical
evolution of the canonical field and matter variables \cite{ADM62,K61,RT74}.

We expand the constraint equations (\ref{ce1}) and (\ref{ce2}) in powers of
$1/c$ where we take into account that (see, e.g., Ref.\ \cite{S95})
\be
\label{ord1}
\begin{array}{c}
m_a\sim{\cal O}\left(\frac{1}{c^2}\right),\quad
\phi\sim{\cal O}\left(\frac{1}{c^2}\right),\quad
{\bf p}_a\sim{\cal O}\left(\frac{1}{c^3}\right),\quad
{\piti ij}\sim{\cal O}\left(\frac{1}{c^3}\right),
\\[2ex]
{\htt ij}\sim{\cal O}\left(\frac{1}{c^4}\right),\quad
{\pitt ij}\sim{\cal O}\left(\frac{1}{c^5}\right).
\end{array}
\ee

To calculate the reduced Hamiltonian (\ref{hred}) up to 3PN order we have to
expand the Hamiltonian constraint equation (\ref{ce1}) up to $1/c^{10}$.
Making use of Eqs.\ (\ref{cc1}), (\ref{cc2}), and (\ref{ord1}), after long
calculations, we obtain
\bea
\label{ce1e}
-\Delta\phi&=&
\sum_a\left[
1-\frac{1}{8}\phi+\frac{1}{64}\phi^2-\frac{1}{512}\phi^3+\frac{1}{4096}\phi^4
+\left(
\frac{1}{2}-\frac{5}{16}\phi+\frac{15}{128}\phi^2-\frac{35}{1024}\phi^3\right)
\frac{{\bf p}_a^2}{m_a^2}
\right.\nonumber\\[2ex]&&
+\left(-\frac{1}{8}+\frac{9}{64}\phi-\frac{45}{512}\phi^2\right)
\frac{({\bf p}_a^2)^2}{m_a^4}
+\left(\frac{1}{16}-\frac{13}{128}\phi\right)\frac{({\bf p}_a^2)^3}{m_a^6}
-\frac{5}{128}\frac{({\bf p}_a^2)^4}{m_a^8}
\nonumber\\[2ex]&&\left.
+\left(-\frac{1}{2}+\frac{9}{16}\phi+\frac{1}{4}\frac{{\bf 
p}_a^2}{m_a^2}\right)
\frac{p_{ai}p_{aj}}{m_a^2}{\htt ij}
-\frac{1}{16}\left({\htt ij}\right)^2
\right]m_a\delta_a
\nonumber\\[2ex]&&
+\left(1+\frac{1}{8}\phi\right)\left({\piti ij}\right)^2
+\left(2+\frac{1}{4}\phi\right){\piti ij}{\pitt ij}
+\left({\pitt ij}\right)^2
\nonumber\\[2ex]&&
+\left[\left(-\frac{1}{2}+\frac{1}{4}\phi-\frac{5}{64}\phi^2\right)\phi_{,{ij}}
+\left(\frac{3}{16}-\frac{15}{128}\phi\right)\phi_{,i}\phi_{,j}
+2{\piti ik}{\piti jk}\right]{\htt ij}
\nonumber\\[2ex]&&
+\left(\frac{1}{4}-\frac{7}{32}\phi\right)\left({\ghtt ijk}\right)^2
+\left(\frac{1}{2}+\frac{1}{16}\phi\right){\ghtt ijk}{\ghtt ikj}
+\Delta\left[
\left(-\frac{1}{2}+\frac{7}{16}\phi\right)\left({\htt ij}\right)^2\right]
\nonumber\\[2ex]&&
-\left[\frac{1}{2}\phi{\htt ij}{\ghtt ikj}
+\frac{1}{4}\phi_{,k}\left({\htt ij}\right)^2\right]_{,k}
+{\cal O}\left(\frac{1}{c^{12}}\right).
\eea

We also need to expand the momentum constraint equations (\ref{ce2}) up to
$1/c^7$. Using Eqs.\ (\ref{cc1}), (\ref{cc2}), and (\ref{ord1}), we get
\bea
\label{ce2e}
{\piti ij}_{,j}&=&
\left(-\frac{1}{2}+\frac{1}{4}\phi-\frac{5}{64}\phi^2\right)
\sum_a p_{ai}\delta_a
+\left(-\frac{1}{2}+\frac{1}{16}\phi\right)\phi_{,j}{\piti ij}
\nonumber\\[2ex]&&
-\frac{1}{2}\phi_{,j}{\pitt ij}
-{\piti jk}_{,k}{\htt ij}
+{\piti jk}\left(\frac{1}{2}{\ghtt jki}-{\ghtt ijk}\right)
+{\cal O}\left(\frac{1}{c^{8}}\right).
\eea

All functions entering the right-hand sides of Eqs.\ (\ref{ce1e}) and
(\ref{ce2e}) can be written as sums of terms of different orders in $1/c$. To
the orders needed in our calculations, they read
\bea
\label{ord21}
\phi&=&\phi_{(2)}+\phi_{(4)}+\phi_{(6)}+\phi_{(8)}
+{\cal O}\left(\frac{1}{c^{9}}\right),
\\[2ex]
{\piti ij}&=&{\pitiii ij}+{\pitv ij}+{\pitvii ij}
+{\cal O}\left(\frac{1}{c^{8}}\right),
\\[2ex]
{\htt ij}&=&{\httiv ij}+{\httv ij}+{\httvi ij}
+{\cal O}\left(\frac{1}{c^{7}}\right),
\\[2ex]
\label{ord24}
{\pitt ij}&=&{\pittv ij}+{\cal O}\left(\frac{1}{c^{6}}\right).
\eea

Using Eqs.\ (\ref{ord21})--(\ref{ord24}) we extract from Eq.\ (\ref{ce1e}) the
Hamiltonian constraint equations valid at individual even orders in $1/c$, up
to the order $1/c^8$. They are
\bea
\label{lapphi2}
-\Delta\phi_{(2)}&=&\sum_a m_a\delta_a,
\\[2ex]
\label{lapphi4}
-\Delta\phi_{(4)}&=&\sum_a
\left(-\frac{1}{8}\phi_{(2)}+\frac{1}{2}\frac{{\bf p}_a^2}{m_a^2}\right)
m_a\delta_a,
\\[2ex]
\label{lapphi6}
-\Delta\phi_{(6)}&=&\sum_a\left(
\frac{1}{64}\phi_{(2)}^2-\frac{1}{8}\phi_{(4)}
-\frac{5}{16}\phi_{(2)}\frac{{\bf p}_a^2}{m_a^2}
-\frac{1}{8}\frac{({\bf p}_a^2)^2}{m_a^4}\right)m_a\delta_a
\nonumber\\[2ex]&&
+\left({\pitiii ij}\right)^2
-\frac{1}{2}\phi_{(2),ij}{\httiv ij},
\\[2ex]
\label{lapphi8}
-\Delta\phi_{(8)}&=&\sum_a\left[
-\frac{1}{512}\phi_{(2)}^3+\frac{1}{32}\phi_{(2)}\phi_{(4)}
-\frac{1}{8}\phi_{(6)}
+\left(\frac{15}{128}\phi_{(2)}^2-\frac{5}{16}\phi_{(4)}\right)
\frac{{\bf p}_a^2}{m_a^2}
\right.\nonumber\\[2ex]&&\left.
+\frac{9}{64}\phi_{(2)}\frac{({\bf p}_a^2)^2}{m_a^4}
+\frac{1}{16}\frac{({\bf p}_a^2)^3}{m_a^6}
-\frac{1}{2}\frac{p_{ai}p_{aj}}{m_a^2}{\httiv ij}
\right]m_a\delta_a
\nonumber\\[2ex]&&
+\frac{1}{8}\phi_{(2)}\left({\pitiii ij}\right)^2
+2{\pitiii ij}{\pitv ij}+2{\pitiii ij}{\pittv ij}
\nonumber\\[2ex]&&
+\left(\frac{3}{16}\phi_{(2),i}\phi_{(2),j}
+\frac{1}{4}\phi_{(2)}\phi_{(2),ij}-\frac{1}{2}\phi_{(4),ij}\right){\httiv ij}
\nonumber\\[2ex]&&
-\frac{1}{2}\Delta\left({\httiv ij}\right)^2
+\frac{1}{4}\left({\ghttiv ijk}\right)^2
+\frac{1}{2}{\ghttiv ijk}{\ghttiv ikj}
-\frac{1}{2}\phi_{(2),ij}{\httvi ij}.
\eea

Explicit solutions of the equations (\ref{lapphi2}), (\ref{lapphi4}), and
(\ref{lapphi6}) for the functions $\phi_{(2)}$, $\phi_{(4)}$, and $\phi_{(6)}$,
respectively, are shown in Appendix A. The full solution of the equation
(\ref{lapphi8}) for the function $\phi_{(8)}$ is not  known. We split
$\phi_{(8)}$ into two parts
\be
\label{phi8sum}
\phi_{(8)}=\phi_{(8)1}+\phi_{(8)2}.
\ee
The function $\phi_{(8)1}$ is explicitly calculable and can be found in
Appendix A.  The unknown part of $\phi_{(8)}$ we call $\phi_{(8)2}$. It is
given by
\be
\label{phi82def}
\phi_{(8)2}=\Delta^{-1}S_{(8)2},
\ee
where
\bea
\label{S82def}
S_{(8)2}&:=&
-\frac{1}{8}\phi_{(2)}\left({\pitiii ij}\right)^2
-2{\pitiii ij}{\pitv ij}
-2{\pitiii ij}{\pittv ij}
\nonumber\\[2ex]&&
+\left(-\frac{3}{16}\phi_{(2),i}\phi_{(2),j}
-\frac{1}{4}\phi_{(2)}\phi_{(2),ij}\right){\httiv ij}
\nonumber\\[2ex]&&
-\frac{1}{4}\left({\ghttiv ijk}\right)^2
-\frac{1}{2}{\ghttiv ijk}{\ghttiv ikj}
+\frac{1}{2}\phi_{(2),ij}{\httvi ij}.
\eea

Using the Eqs.\ (\ref{ord21})--(\ref{ord24}) we extract from Eq.\ (\ref{ce2e})
the momentum constraint equations valid up to the order $1/c^7$: 
\bea
\label{eqpi3}
{\gpitiii ijj}&=&-\frac{1}{2}\sum_a p_{ai}\delta_a,
\\[2ex]
\label{eqpi5}
{\gpitv ijj}&=&-\frac{1}{2}\left(\phi_{(2)}{\pitiii ij}\right)_{,j},
\\[2ex]
\label{eqpi71}
{\gpitvii ijj}&=&\mit\Gamma^{ij}_{(7),j},
\eea
where
\bea
\label{eqpi72}
\mit\Gamma^{ij}_{(7)}&:=&
-\frac{1}{2}\left[
\phi_{(2)}{\pitv ij}
+\left(\frac{3}{16}\phi_{(2)}^2+\phi_{(4)}\right){\pitiii ij}
+\phi_{(2)}{\pittv ij}
+2{\pitiii jk}{\httiv ik}
\right.\nonumber\\[2ex]&&\left.
-\left(2\pi_{(3)}^k+\Delta^{-1}\pi_{(3),kl}^l\right){\ghttiv jki}\right].
\eea

Explicit solutions of the equations (\ref{eqpi3}) and (\ref{eqpi5}) for the
functions ${\pitiii ij}$ and ${\pitv ij}$, respectively,  are given in Appendix
A, where one can also find the formula for the function $\pi_{(3)}^i$ connected
with ${\pitiii ij}$ by means of Eq.\ (\ref{pitidec}).

The TT-part of the metric in the leading order fulfils the equation (see, e.g.,
Eq.\ (14) of  \cite{JS97})
\be
\label{httiv}
\Delta{\httiv ij}=\delta_{ij}^{{\rm TT}kl}A_{(4)kl},
\ee
where
\be
\label{r6a}
A_{(4)ij}
:= -\sum_a\frac{p_{ai}p_{aj}}{m_a}\delta_a
-\frac{1}{4}\phi_{(2),i}\phi_{(2),j},
\ee
and where $\delta_{ij}^{{\rm TT}kl}$ is the TT-projection operator defined by
(see,  e.g., Eq.\ (4) of \cite{JS97})
\bea
\label{r6c}
\delta^{{\rm TT}kl}_{ij} &:=&
\frac{1}{2}\left[
(\delta_{il}-\Delta^{-1}\pa_i\pa_l) (\delta_{jk}-\Delta^{-1}\pa_j\pa_k)
+ (\delta_{ik}-\Delta^{-1}\pa_i\pa_k) (\delta_{jl}-\Delta^{-1}\pa_j\pa_l)
\right.\nonumber\\&&
- (\delta_{kl}-\Delta^{-1}\pa_k\pa_l) (\delta_{ij}-\Delta^{-1}\pa_i\pa_j)
\left.\right].
\eea
The explicit formula for the function ${\httiv ij}$ can be found in Appendix A.

In our calculations we also need the explicit formula for that part of the
metric function ${\httvi ij}$ which diverges linearly at infinity. So we write
\be
\label{htt6a}
{\httvi ij} = h_{(6)ij}^{\rm TT div} + h_{(6)ij}^{\rm TT conv},
\ee
where $h_{(6)ij}^{\rm TT div}\sim r$ and $h_{(6)ij}^{\rm TT conv}\sim 1/r$ as
$r\to\infty$. The function $h_{(6)ij}^{\rm TT div}$ is given by the integral
(see, e.g., Eq.\ (12) in \cite{JS97})
\be
\label{htt6b}
h_{(6)ij}^{\rm TT div}\left({\bf x},t\right)
= -\frac{1}{8\pi} \delta_{ij}^{{\rm TT}kl}
\myintb \frac{\pa^2 A_{(4)kl}}{\pa t^2}\left({\bf x}',t\right)
\mid{\bf x}-{\bf x}'\mid
= \delta_{ij}^{{\rm TT}kl}\frac{\pa^2}{\pa t^2}
\left(\Delta^{-2}A_{(4)kl}\right)\left({\bf x},t\right),
\ee
where the function $A_{(4)ij}$ is defined in Eq.\ (\ref{r6a}). We have
calulated the integral from the right-hand side of Eq.\ (\ref{htt6b}). The
result is given in Appendix A.

\section{The conservative Hamilton function up to 3PN order}

In the following we are interested in a {\em conservative} 3PN Hamiltonian
which depends on body variables only. To achieve this goal we firstly
transform our field-plus-matter Hamilton functional (\ref{hred}) into a Routh
functional which is a Hamilton function for the bodies but a Lagrange
functional for the field (notice for the following the crucial difference
between a Hamiltonian and a Lagrangian: the functional derivative of the
latter is zero, the one of the former not). The Routh functional up to 3PN
order turns out to be
\be
\label{rou1}
R_{\le 3{\rm PN}}
\left[{\bf x}_a,{\bf p}_{a},{\httiv ij},{\httivdot ij}\right]
= H_{\le 3{\rm PN}} - \myinta{\pittv ij}{\httivdot ij},
\ee
where $H_{\le 3{\rm PN}}$ is the Hamiltonian up to 3PN order. The equations of
motion of the point masses determined by the Routh functional (\ref{rou1}) read
\be
\label{rou3}
\dot{\bf p}_a = -\frac{\pa R_{\le 3{\rm PN}}}{\pa{\bf x}_a},\quad
\dot{\bf x}_a = \frac{\pa R_{\le 3{\rm PN}}}{\pa{\bf p}_a}.
\ee
We eliminate now in the Routh functional (\ref{rou1}) ${\httiv ij}$ and
${\httivdot ij}$ by ${\bf x}_a$, $\dot{\bf x}_a$, ${\bf p}_a$, and $\dot{\bf
p}_a$ through solving the field equations which result from
$H_{\le 3{\rm PN}}$ (see Eq.\ (\ref{httiv})).
After that the Routh functional (\ref{rou1}) becomes a higher order
matter Hamilton function (denoted by a tilde) of the variables ${\bf x}_a$,
$\dot{\bf x}_a$, ${\bf p}_a$, and $\dot{\bf p}_a$:
\be
\label{rou4}
\widetilde{H}_{\le 3{\rm PN}}
\left({\bf x}_a,{\bf p}_a,\dot{\bf x}_a,\dot{\bf p}_a\right)
= R_{\le 3{\rm PN}}
\left[
{\bf x}_a,
{\bf p}_{a},
{\httiv ij}\left({\bf x}_a,{\bf p}_a\right),
{\httivdot ij}\left({\bf x}_a,{\bf p}_a,\dot{\bf x}_a,\dot{\bf p}_a\right)
\right].
\ee
The equations of motion of the bodies determined by the Hamilton function
(\ref{rou4}) read
\be
\dot{\bf p}_a = -\frac{\pa \widetilde{H}_{\le 3{\rm PN}}}{\pa{\bf x}_a}
+\frac{d}{dt}
\left(\frac{\pa \widetilde{H}_{\le 3{\rm PN}}}{\pa\dot{\bf x}_a}\right),\quad
\dot{\bf x}_a = \frac{\pa \widetilde{H}_{\le 3{\rm PN}}}{\pa{\bf p}_a}
-\frac{d}{dt}
\left(\frac{\pa \widetilde{H}_{\le 3{\rm PN}}}{\pa\dot{\bf p}_a}\right).
\ee
In these equations the higher time derivatives may be eliminated by applying
lower order equations of motion. The elimination of the higher derivatives
in the Hamilton function (\ref{rou4}) would result in a redefinition of the body
variables (see, e.g., Ref.\ \cite{DS91}).

The reduced Hamiltonian $H_{\le 3{\rm PN}}$ of Eq.\ (\ref{rou1}) can be
obtained from Eq.\ (\ref{ce1e}).  After dropping full divergences (including
the term ${\piti ij}{\pitt ij}$, which can be written as
$[(2\pi^j+\Delta^{-1}\pi^k_{,jk}){\pitt ij}]_{,i}$, cf.\ Eq.\
(\ref{pitidec})), it reads
\bea
\label{hame}
H_{\le 3{\rm PN}} &=& \myinta\left\{ \sum_a\left[
1-\frac{1}{8}\phi+\frac{1}{64}\phi^2-\frac{1}{512}\phi^3+\frac{1}{4096}\phi^4
\right.\right.\nonumber\\[2ex]&&
+\left(
\frac{1}{2}-\frac{5}{16}\phi+\frac{15}{128}\phi^2-\frac{35}{1024}\phi^3\right)
\frac{{\bf p}_a^2}{m_a^2}
\nonumber\\[2ex]&&
+\left(-\frac{1}{8}+\frac{9}{64}\phi-\frac{45}{512}\phi^2\right)
\frac{({\bf p}_a^2)^2}{m_a^4}
+\left(\frac{1}{16}-\frac{13}{128}\phi\right)\frac{({\bf p}_a^2)^3}{m_a^6}
-\frac{5}{128}\frac{({\bf p}_a^2)^4}{m_a^8}
\nonumber\\[2ex]&&\left.
+\left(-\frac{1}{2}+\frac{9}{16}\phi+\frac{1}{4}\frac{{\bf 
p}_a^2}{m_a^2}\right)
\frac{p_{ai}p_{aj}}{m_a^2}{\htt ij}
-\frac{1}{16}\left({\htt ij}\right)^2
\right]m_a\delta_a
\nonumber\\[2ex]&&
+\left(1+\frac{1}{8}\phi\right)\left({\piti ij}\right)^2
+\frac{1}{4}\phi{\piti ij}{\pitt ij}
+\left({\pitt ij}\right)^2
\nonumber\\[2ex]&&
+\left[\left(\frac{1}{4}\phi-\frac{5}{64}\phi^2\right)\phi_{,{ij}}
+\left(\frac{3}{16}-\frac{15}{128}\phi\right)\phi_{,i}\phi_{,j}
+2{\piti ik}{\piti jk}\right]{\htt ij}
\nonumber\\[2ex]&&\left.
+\left(\frac{1}{4}-\frac{7}{32}\phi\right)\left({\ghtt ijk}\right)^2
+\left(\frac{1}{16}\phi\right){\ghtt ijk}{\ghtt ikj} \right\}.
\eea
Le us note that the integrals of the full divergences we have dropped in Eq.\
(\ref{hame}) do not contribute to the Hamiltonian because they fall off at
infinity at least as $1/r^4$. This can be inferred from the following asymptotic
behaviour of the functions entering on the right-hand side of Eq.\ (\ref{ce1e})
(cf.\ Ref.\ \cite{RT74}):
\be
\phi\sim\frac{1}{r},\quad
{\htt ij}\sim\frac{1}{r},\quad
{\piti ij}\sim\frac{1}{r^2},\quad
{\pitt ij}\sim\frac{1}{r^2},\qquad {\rm for}\quad r\to\infty;
\ee
see also our discussion below Eq.\ (\ref{r4reg}).

The higher order Hamiltonian $\widetilde{H}_{\le 3{\rm PN}}$
of Eq.\ (\ref{rou4}) can be split as follows
\bea
\widetilde{H}_{\le 3{\rm PN}}
\left({\bf x}_a,{\bf p}_a,\dot{\bf x}_a,\dot{\bf p}_a\right)
&=&\widetilde{H}_0
+\widetilde{H}_{\rm N}\left({\bf x}_a,{\bf p}_a\right)
+\widetilde{H}_{\rm 1PN}\left({\bf x}_a,{\bf p}_a\right)
\nonumber\\[2ex]&&
+\widetilde{H}_{\rm 2PN}\left({\bf x}_a,{\bf p}_a\right)
+\widetilde{H}_{\rm 3PN}
\left({\bf x}_a,{\bf p}_a,\dot{\bf x}_a,\dot{\bf p}_a\right),
\eea
where the  Hamiltonians $\widetilde{H}_0$ through $\widetilde{H}_{\rm 3PN}$
can be extracted from Eqs.\ (\ref{rou1}) and (\ref{hame}) by
means of the expansions (\ref{ord21})--(\ref{ord24}). The Hamiltonians
$\widetilde{H}_0$ through $\widetilde{H}_{\rm 2PN}$ are known. We re-calculate
them below for completeness. Our aim, however, is to calculate the 3PN
Hamilton function $\widetilde{H}_{\rm 3PN}$.

The calculation of the Hamiltonians $\widetilde{H}_0$ through
$\widetilde{H}_{\rm 2PN}$ can be performed directly as we explicitly know all
functions needed to perform the integrations. The Hamiltonians can be written
in the form
\bea
\label{h0def}
\widetilde{H}_0
&=& \myinta \sum_a m_a\delta_a,
\\[2ex]
\label{hndef}
\widetilde{H}_{\rm N}
&=& \myinta \sum_a
\left(-\frac{1}{8}\phi_{(2)}+\frac{1}{2}\frac{{\bf p}_a^2}{m_a^2}\right)
m_a\delta_a,
\\[2ex]
\widetilde{H}_{\rm 1PN} &=& \widetilde{H}_{11} + \widetilde{H}_{12},
\\[2ex]
\label{h11def}
\widetilde{H}_{11}
&=& \myinta \sum_a\left(
\frac{1}{64}\phi_{(2)}^2-\frac{1}{8}\phi_{(4)}
-\frac{5}{16}\phi_{(2)}\frac{{\bf p}_a^2}{m_a^2}
-\frac{1}{8}\frac{({\bf p}_a^2)^2}{m_a^4}\right)m_a\delta_a,
\\[2ex]
\label{h12def}
\widetilde{H}_{12}
&=& \myinta \left\{ \left({\pitiii ij}\right)^2 \right\},
\eea
\bea
\widetilde{H}_{\rm 2PN}
&=& \widetilde{H}_{21} + \widetilde{H}_{22} + \widetilde{H}_{23},
\\[2ex]
\label{h21def}
\widetilde{H}_{21}
&=& \myinta \sum_a\left[
-\frac{1}{512}\phi_{(2)}^3+\frac{1}{32}\phi_{(2)}\phi_{(4)}
-\frac{1}{8}\phi_{(6)}
+\left(\frac{15}{128}\phi_{(2)}^2-\frac{5}{16}\phi_{(4)}\right)
\frac{{\bf p}_a^2}{m_a^2}
\right.\nonumber\\[2ex]&&\left.
+\frac{9}{64}\phi_{(2)}\frac{({\bf p}_a^2)^2}{m_a^4}
+\frac{1}{16}\frac{({\bf p}_a^2)^3}{m_a^6}
-\frac{1}{2}\frac{p_{ai}p_{aj}}{m_a^2}{\httiv ij}
\right]m_a\delta_a,
\\[2ex]
\label{h22def}
\widetilde{H}_{22}
&=& \myinta \left\{
\frac{1}{8}\phi_{(2)}\left({\pitiii ij}\right)^2
+2{\pitiii ij}{\pitv ij}
-\frac{1}{16}\phi_{(2),i}\phi_{(2),j}{\httiv ij}
+\frac{1}{4}\left({\ghttiv ijk}\right)^2 \right\},
\\[2ex]
\label{h23def}
\widetilde{H}_{23}
&=& \myinta \left\{
\frac{1}{4}\phi_{(2)}\phi_{(2),j}{\httiv ij} \right\}_{,i}.
\eea

The Hamiltonian $\widetilde{H}_{\rm 3PN}$, as extracted from Eqs.\
(\ref{rou1}) and (\ref{hame}), reads
\bea
\label{h3pn}
\widetilde{H}_{\rm 3PN}
&=& \myinta\left\{
\sum_a\left[
\frac{1}{4096}\phi_{(2)}^4-\frac{3}{512}\phi_{(2)}^2\phi_{(4)}
+\frac{1}{64}\phi_{(4)}^2+\frac{1}{32}\phi_{(2)}\phi_{(6)}
-\frac{1}{8}\phi_{(8)}
\right.\right.\nonumber\\[2ex]&&
+\left(-\frac{35}{1024}\phi_{(2)}^3
+\frac{15}{64}\phi_{(2)}\phi_{(4)}
-\frac{5}{16}\phi_{(6)}\right)\frac{{\bf p}_a^2}{m_a^2}
\nonumber\\[2ex]&&
+\left(-\frac{45}{512}\phi_{(2)}^2+\frac{9}{64}\phi_{(4)}\right)
\frac{({\bf p}_a^2)^2}{m_a^4}
-\frac{13}{128}\phi_{(2)}\frac{({\bf p}_a^2)^3}{m_a^6}
-\frac{5}{128}\frac{({\bf p}_a^2)^4}{m_a^8}
\nonumber\\[2ex]&&\left.
+\left(\frac{9}{16}\phi_{(2)}+\frac{1}{4}\frac{{\bf p}_a^2}{m_a^2}\right)
\frac{p_{ai}p_{aj}}{m_a^2}{\httiv ij}
-\frac{1}{16}\left({\httiv ij}\right)^2
\right]m_a\delta_a
\nonumber\\[2ex]&&
+\frac{1}{8}\phi_{(4)}\left({\pitiii ij}\right)^2
+\frac{1}{4}\phi_{(2)}{\pitiii ij}{\pitv ij}
+\left({\pitv ij}\right)^2
+2{\pitiii ij}{\pitvii ij}
+\left({\pittv ij}\right)^2
\nonumber\\[2ex]&&
+\frac{1}{4}\phi_{(2)}{\pitiii ij}{\pittv ij}
+2{\pitiii ik}{\pitiii jk}{\httiv ij}
+\left(-\frac{15}{128}\phi_{(2)}\phi_{(2),i}\phi_{(2),j}
-\frac{5}{64}\phi_{(2)}^2\phi_{(2),ij}
\right.\nonumber\\[2ex]&&\left.
+\frac{1}{4}\phi_{(2)}\phi_{(4),ij}
+\frac{3}{8}\phi_{(2),i}\phi_{(4),j}
+\frac{1}{4}\phi_{(2),ij}\phi_{(4)}\right){\httiv ij}
-\frac{7}{32}\phi_{(2)}\left({\ghttiv ijk}\right)^2
\nonumber\\[2ex]&&
+\frac{1}{16}\phi_{(2)}{\ghttiv ijk}{\ghttiv ikj}
-\frac{1}{2}\sum_a\frac{p_{ai}p_{aj}}{m_a}{\httvi ij}\delta_a
+\left(\frac{3}{16}\phi_{(2),i}\phi_{(2),j}
+\frac{1}{4}\phi_{(2)}\phi_{(2),ij}\right){\httvi ij}
\nonumber\\[2ex]&&\left.
+\frac{1}{2}{\ghttiv ijk}{\ghttvi ijk}
-{\pittv ij}{\httivdot ij} \right\}.
\eea
For the calculation of the Hamiltonian $\widetilde{H}_{\rm 3PN}$ we perform
some manipulations which allow us to do the integrations in Eq.\ (\ref{h3pn}) 
without explicit knowledge of all functions entering on the right-hand side of
Eq.\ (\ref{h3pn}). There are three such functions: the part $\phi_{(8)2}$ of
the function $\phi_{(8)}$ [cf.\ Eq.\ (\ref{phi8sum})], the function ${\pitvii
ij}$, and the part $h_{(6)ij}^{\rm TT conv}$ of the function ${\httvi ij}$.

The function $\phi_{(8)2}$ we eliminate by means of the identity
\be
\label{phi8div}
\left(\Delta\phi_{(2)}\right)\phi_{(8)2}
=\phi_{(2)}S_{(8)2}
+\left(\phi_{(2),i}\phi_{(8)2}-\phi_{(2)}\phi_{(8)2,i}\right)_{,i}.
\ee

To eliminate the unknown function ${\pitvii ij}$ we use the following relation,
which can be proved by means of the Eqs.\ (\ref{eqpi71}) and (\ref{eqpi72}), and by
the aid of the Eq.\ (\ref{pitidec}):
\bea
\label{pi3pi7}
{\pitiii ij}{\pitvii ij}&=&
\left[\left(2\pi_{(3)}^i+\Delta^{-1}\pi_{(3),ik}^k\right)
\left({\pitvii ij}-\mit\Gamma_{(7)}^{ij}\right)\right]_{,j}
-\frac{3}{32}\phi_{(2)}^2\left({\pitiii ij}\right)^2
-\frac{1}{2}\phi_{(4)}\left({\pitiii ij}\right)^2
\nonumber\\[2ex]&&
-\frac{1}{2}\phi_{(2)}{\pitiii ij}{\pitv ij}
-\frac{1}{2}\phi_{(2)}{\pitiii ij}{\pittv ij}
-\left(2\pi_{(3)}^i+\Delta^{-1}\pi_{(3),il}^l\right)_{,k}
{\pitiii jk}{\httiv ij}
\nonumber\\[2ex]&&
+\frac{1}{2}\left(2\pi_{(3)}^i+\Delta^{-1}\pi_{(3),il}^l\right)
\left(2\pi_{(3),j}^k+\Delta^{-1}\pi_{(3),jkm}^m\right){\ghttiv ijk}.
\eea

All the terms which depend on the unknown function ${\httvi ij}$ we were be
able to write as a full divergence [see Eq.\ (\ref{r6def}) below]. To do this
we have used the following relation, which can be derived using the explicit
formula (\ref{r6c}) for the $\delta_{ij}^{{\rm TT}kl}$ operator and the
traceless property of the function ${\httvi ij}$:
\bea
\label{htt6div}
&\kern-25em \left(\delta_{ij}^{{\rm TT}kl}A_{(4)kl}-A_{(4)ij}\right){\httvi ij}&
\nonumber\\[2ex]
&=  \left\{\left[
\frac{1}{2}\left(\Delta^{-1}A_{(4)kk,j}\right)
-2\left(\Delta^{-1}A_{(4)jk,k}\right)
+\frac{1}{2}\left(\Delta^{-2}A_{(4)kl,jkl}\right)\right]{\httvi ij}
\right\}_{,i}.&
\eea

In the last stage we eliminate the field momentum ${\pittv ij}$ by ${\bf x}_a$,
$\dot{\bf x}_a$, ${\bf p}_a$, and $\dot{\bf p}_a$.  We use the field equation
for the field momentum in the leading order. It reads (see, e.g., Eq.\ (13) of
\cite{JS97}):
\be
\label{pitt51}
{\pittv ij}
=\frac{1}{2}{\httivdot ij}
+\frac{1}{2}\left(\phi_{(2)}{\pitiii ij}\right)^{\rm TT}.
\ee
One can check that the following relation holds
\be
\label{pitt52}
\left(\phi_{(2)}{\pitiii ij}\right)^{\rm TT}=
\phi_{(2)}{\pitiii ij}+2{\pitv ij}.
\ee
Substituting Eq.\ (\ref{pitt52}) into Eq.\ (\ref{pitt51}) we obtain
\be
\label{pitt53}
{\pittv ij}
=\frac{1}{2}{\httivdot ij}
+\frac{1}{2}\phi_{(2)}{\pitiii ij}
+{\pitv ij}.
\ee

Using Eqs.\ (\ref{phi8sum}), (\ref{S82def}), (\ref{phi8div}), (\ref{pi3pi7}),
(\ref{htt6div}), and (\ref{pitt53}) we rewrite the Hamilton function 
$\widetilde{H}_{\rm 3PN}$ given by Eq.\ (\ref{h3pn}) as a sum of terms
\be
\label{r3sum}
\widetilde{H}_{\rm 3PN}=\sum_{I=1}^6 \widetilde{H}_{3I},
\ee
where the $\widetilde{H}_{3I}$ are defined as follows
\bea
\label{r1def}
\widetilde{H}_{31}&:=&\myinta\sum_a\left\{
\frac{1}{4096}\phi_{(2)}^4-\frac{3}{512}\phi_{(2)}^2\phi_{(4)}
+\frac{1}{64}\phi_{(4)}^2+\frac{1}{32}\phi_{(2)}\phi_{(6)}
-\frac{1}{8}\phi_{(8)1}
\right.\nonumber\\[2ex]&&\kern-6ex
+\left(-\frac{35}{1024}\phi_{(2)}^3
+\frac{15}{64}\phi_{(2)}\phi_{(4)}
-\frac{5}{16}\phi_{(6)}\right)\frac{{\bf p}_a^2}{m_a^2}
+\left(-\frac{45}{512}\phi_{(2)}^2+\frac{9}{64}\phi_{(4)}\right)
\frac{({\bf p}_a^2)^2}{m_a^4}
\nonumber\\[2ex]&&\left.\kern-6ex
-\frac{13}{128}\phi_{(2)}\frac{({\bf p}_a^2)^3}{m_a^6}
-\frac{5}{128}\frac{({\bf p}_a^2)^4}{m_a^8}
+\left(\frac{9}{16}\phi_{(2)}+\frac{1}{4}\frac{{\bf p}_a^2}{m_a^2}\right)
\frac{p_{ai}p_{aj}}{m_a^2}{\httiv ij}
-\frac{1}{16}\left({\httiv ij}\right)^2
\right\}m_a\delta_a,
\\[2ex]
\label{r2def}
\widetilde{H}_{32}&:=&\myinta\left\{
-\frac{7}{8}\phi_{(4)}\left({\pitiii ij}\right)^2
-\frac{1}{8}\phi_{(2),i}\phi_{(4),j}{\httiv ij}
-\left(\left(\phi_{(2)}{\pitiii ij}\right)^{\rm TT}\right)^2\right\},
\\[2ex]
\label{r3def}
\widetilde{H}_{33}&:=&\myinta\left\{
-\frac{1}{4}\left({\httivdot ij}\right)^2
-\frac{1}{2}{\httivdot ij}\left(\phi_{(2)}{\pitiii ij}\right)^{\rm TT}
\right\},
\\[2ex]
\label{r4def}
\widetilde{H}_{34}&:=&\myinta\left\{
\frac{35}{64}\phi_{(2)}^2\left({\pitiii ij}\right)^2
+2{\pitiii ik}{\pitiii jk}{\httiv ij}
-2\left(2\pi_{(3)}^i+\Delta^{-1}\pi_{(3),il}^l\right)_{,k}
{\pitiii jk}{\httiv ij}
\right.\nonumber\\[2ex]&&
+\left(2\pi_{(3)}^i+\Delta^{-1}\pi_{(3),il}^l\right)
\left(2\pi_{(3)}^k+\Delta^{-1}\pi_{(3),km}^m\right)_{,j}{\ghttiv ijk}
\nonumber\\[2ex]&&\left.
+\frac{5}{64}\phi_{(2)}\phi_{(2),i}\phi_{(2),j}{\httiv ij}
-\frac{1}{4}\phi_{(2)}\left({\ghttiv ijk}\right)^2\right\},
\\[2ex]
\label{r5def}
\widetilde{H}_{35}&:=&\myinta\left\{
-\frac{7}{64}\phi_{(2)}^2\phi_{(2),j}{\httiv ij}
+\frac{1}{4}\left(\phi_{(2)}\phi_{(4)}\right)_{,j}{\httiv ij}
\right.\nonumber\\[2ex]&&
+\left(2\pi_{(5)}^j+\Delta^{-1}\pi_{(5),jk}^k\right)
\left({\httivdot ij}+3\left(\phi_{(2)}{\pitiii ij}\right)^{\rm TT}\right)
\nonumber\\[2ex]&&\left.
+2\left(2\pi_{(3)}^j+\Delta^{-1}\pi_{(3),jk}^k\right)
\left({\pitvii ji}-\mit\Gamma_{(7)}^{ji}\right)
+\frac{1}{8}\left(\phi_{(2),i}\phi_{(8)2}-\phi_{(2)}\phi_{(8)2,i}\right)
\right\}_{,i},
\\[2ex]
\label{r6def}
\widetilde{H}_{36}
&:=&\myinta \left\{
\frac{5}{16}\phi_{(2)}\phi_{(2),j}{\httvi ij}
+\frac{1}{2}{\ghttiv jki}{\httvi jk}
\right.\nonumber\\[2ex]&&\left.
-\frac{1}{2}\left[
\frac{1}{2}\left(\Delta^{-1}A_{(4)kk,j}\right)
-2\left(\Delta^{-1}A_{(4)jk,k}\right)
+\frac{1}{2}\left(\Delta^{-2}A_{(4)kl,jkl}\right)\right]{\httvi ij}
\right\}_{,i}.
\eea

\section{Results of regularization procedures}

To diminish the number of terms we perform calculations in the center-of-mass 
reference frame, so we can use the relations
\be
\label{cm}
{\bf p}_1+{\bf p}_2=0,\quad
\dot{\bf p}_1+\dot{\bf p}_2=0.
\ee
We also use, at the 3PN level, the Newtonian relations between the coordinate velocities
$\dot{\bf x}_1$, $\dot{\bf x}_2$ of the bodies and their relative velocity 
${\bf v}:=\dot{\bf x}_1-\dot{\bf x}_2$:
\be
\label{v1v2}
\dot{\bf x}_1=\frac{m_2}{M}{\bf v},\quad
\dot{\bf x}_2=-\frac{m_1}{M}{\bf v},
\ee
where $M:=m_1+m_2$ is the total mass of the system.

To shorten the formulae we introduce the following reduced variables
\be
{\bf r}:=\frac{16\pi}{M}\left({\bf x}_1-{\bf x}_2\right),\quad
r := |{\bf r}|,\quad{\bf n}:=\frac{\bf r}{r},\quad
{\bf p}:=\frac{{\bf p}_1}{\mu}=-\frac{{\bf p}_2}{\mu},\quad
{\bf q}:=\frac{\dot{\bf p}_1}{16\pi\nu}=-\frac{\dot{\bf p}_2}{16\pi\nu},
\ee
where
\be
\mu:=\frac{m_1 m_2}{M},\quad
\nu:=\frac{\mu}{M}.
\ee
We also introduce the reduced Hamilton function
\be
\widehat{H}:=\frac{\widetilde{H}}{\mu}.
\ee
$\widehat{H}$ depends on masses of the binary system only through the parameter
$\nu$. From now on the hat will indicate division by the reduced mass $\mu$.

To calculate the Hamiltonians $\widetilde{H}_0$ through $\widetilde{H}_{\rm
2PN}$ we proceed as follows. The terms $\widetilde{H}_0$, $\widetilde{H}_{\rm
N}$, $\widetilde{H}_{11}$, and $\widetilde{H}_{21}$ [given by Eqs.\
(\ref{h0def}), (\ref{hndef}), (\ref{h11def}), and (\ref{h21def}), respectively]
we regularize by means of the Hadamard's procedure described in Appendix B.1.
The terms $\widetilde{H}_{12}$ and $\widetilde{H}_{22}$ [given by Eqs.\
(\ref{h12def}) and (\ref{h22def}), respectively] we regularize using the
procedure from Appendix B.2. The term $\widetilde{H}_{23}$ [from Eq.\
(\ref{h23def})] is a full divergence. The integrand in $\widetilde{H}_{23}$
falls off at infinity as $1/r^4$ so $\widetilde{H}_{23}$ does not contribute to
the Hamiltonian. We have checked that the regularization procedure of Appendix
B.2 applied to $\widetilde{H}_{23}$ gives zero. The final result for the
Hamiltonian $\widetilde{H}_0 + \widetilde{H}_{\rm N} + \widetilde{H}_{\rm 1PN}
+ \widetilde{H}_{\rm 2PN}$ coincides with that known in the literature (see,
e.g., Eq.\ (3.1) in \cite{DS88}).

The calculation of the Hamiltonian $\widetilde{H}_{\rm 3PN}$ is much more
complicated. The term $\widetilde{H}_{31}$ of Eq.\ (\ref{r1def}) we regularize
using the Hadamard's procedure described in Appendix B.1. After long
calculations we obtain
\bea
\label{r1reg}
\kern-4em\widehat{H}_{31}
&=&\frac{1}{128}\left(-5+35\nu-70\nu^2+35\nu^3\right)\ppp^4
\nonumber\\&&\kern-3em
+\frac{1}{16}\bigg\{
\left(-7+42\nu-53\nu^2-6\nu^3\right)\ppp^3
+(1-2\nu)\nu^2\left[2\np^2\ppp^2+3\np^4\pp\right]
\bigg\}\frac{1}{r}
\nonumber\\&&\kern-3em
+\frac{1}{64}\bigg\{
\left(-108+515\nu+84\nu^2\right)\ppp^2
+(46+35\nu)\nu\np^2\pp-(45-255\nu)\nu\np^4
\bigg\}\frac{1}{r^2}
\nonumber\\&&\kern-3em
+\frac{1}{160}\bigg\{
\left(-500+1760\nu-2317\nu^2\right)\pp-3(480-1897\nu)\nu\np^2
\bigg\}\frac{1}{r^3}
+\frac{1}{16}(2-11\nu)\frac{1}{r^4}.
\eea

The calculation of $\widetilde{H}_{32}$, $\widetilde{H}_{33}$, and
$\widetilde{H}_{34}$ [defined by Eqs.\ (\ref{r2def})--(\ref{r4def})] we do by
means of the regularization procedure described in Appendix B.2.  The
performance of the procedure differs  considerably in the case of
$\widetilde{H}_{32}$ and $\widetilde{H}_{33}$ compared to
$\widetilde{H}_{34}$.  In $\widetilde{H}_{32}$ and $\widetilde{H}_{33}$ each
term can be regularized separately.  This is not the case for
$\widetilde{H}_{34}$.  For each term of $\widetilde{H}_{34}$ taken separately
the limit (\ref{r03}) of Appendix B.2 does not exist.  By series expansion with
respect to $\ve$ one can see that for each term entering $\widetilde{H}_{34}$
the sum from Eq.\ (\ref{r03}) contains a part proportional to $1/\ve$ (similar
divergences arise in the dimensional regularization procedure in quantum field
theory, see, e.g., \cite{IZ80}). After collecting {\em all} $1/\ve$ terms (they
are of the type (\ref{amb2}) only) the  parts cancel each other. Furthermore,
after regularization, there are terms of the type (\ref{amb2}) proportional to
$\ln r_{12}$.  After collecting all such terms the logarithms cancel out, too.
Unfortunately, the result of application of the regularization procedure to
$\widetilde{H}_{34}$ is not unique. The ambiguity which arises can be expressed
in terms of exactly one unknown number and will be discussed in detail in the
next section.

The result of applying the regularization procedure of Appendix B.2 to Eqs.\
(\ref{r2def})--(\ref{r4def}), after tedious calculations, is
\bea
\label{r2reg}
\widehat{H}_{32}&=&
\frac{3}{64}\,\nu\,\bigg\{
(-4+113\nu)\ppp^2+(12-31\nu)\np^2\pp
\bigg\}\frac{1}{r^2}
\nonumber\\&&
+\frac{1}{512}\,\nu\,\bigg\{
\left[603\pi^2-9632-4\left(1004+15\pi^2\right)\nu\right]\pp
\nonumber\\&&
+3\left[4640-603\pi^2+12\left(5\pi^2-44\right)\nu\right]\np^2
\bigg\}\frac{1}{r^3}
+\frac{1}{4}\,\nu\,\frac{1}{r^4},
\eea
\bea
\label{r3reg}
\widehat{H}_{33}
&=&\frac{1}{12}\,\nu^2\,\bigg\{
4\np\ndp\pdp
-5\ndp^2\pp
\nonumber\\&&
-5\np^2\dpdp
-\np^2\ndp^2
+13\pp\dpdp
+2\pdp^2
\bigg\}r
\nonumber\\&&
+ \frac{1}{8}\, \nu^2\, \bigg\{
\ndr\pdp \Big[ 5\pp + \np^2 \Big]
- \np \Big[\pp\dpdr+2\pdp\pdr\Big]
\nonumber\\&&
- \frac{1}{3} \np^3\dpdr
+ \ndp \Big[\pp+\np^2\Big] \Big[\np\ndr-\pdr\Big]
\bigg\}
\nonumber\\&&
+\frac{1}{48}\,\nu\,\bigg\{
4(17-10\nu)\np\ndp\ndr
-2(15+22\nu)\np^2\ndp
\nonumber\\&&
-2(51-8\nu)\ndp\pp
-4(6-5\nu)\np\pdp
-4(1-2\nu)\ndr\pdp
\nonumber\\[1ex]&&
-4(7-2\nu)\Big[\np\dpdr+\ndp\pdr\Big]
+3\,\nu^2\,\Big[
 8\np\ndr\pp\pdr
\nonumber\\[1ex]&&
+8\np^3\ndr\pdr
+2\np^2\pp\drdr 
+5\ppp^2\drdr
+\np^4\drdr
-5\ndr^2\ppp^2
\nonumber\\&&
-6\np^2\ndr^2\pp
-5\np^4\ndr^2
-4\np^2\dpdr^2
-4\pp\dpdr^2 \Big]
\bigg\}\frac{1}{r}
\nonumber\\&&
+\frac{1}{48}\,\nu\,\bigg\{
 5(5-7\nu)\np^3\ndr
+10(3-5\nu)\np^2\ndr^2
\nonumber\\&&
+3(17-35\nu)\np\ndr\pp
-28(3-8\nu)\np\ndr\pdr
\nonumber\\[1ex]&&
-15(2-3\nu)\np^2\pdr
+2(24-77\nu)\np^2\drdr
+2(9-29\nu)\ndr^2\pp
\nonumber\\&&
-3(4-9\nu)\pp\pdr
-2(12-37\nu)\pp\drdr
+4(6-17\nu)\pdr^2
\bigg\}\frac{1}{r^2}
\nonumber\\&&
+\frac{1}{1536}\,\nu\,\bigg\{
 3\Big[927\pi^2-10832+36(48-5\pi^2)\nu\Big]\np\ndr
\nonumber\\&&
-6\Big[3(16+\pi^2)-2(45\pi^2-464)\nu\Big]\ndr^2
+\Big[11600-927\pi^2+20(9\pi^2-80)\nu\Big]\pdr
\nonumber\\&&
+2\Big[176+3\pi^2+6(176-15\pi^2)\nu\Big]\,\drdr
\bigg\}\frac{1}{r^3},
\eea
\bea
\label{r4reg}
\widehat{H}_{34}
&=&\frac{1}{192}\,\nu\,\bigg\{
9(19-13\nu)\ppp^2-18(3-43\nu)\np^2\pp-(5-303\nu)\np^4
\bigg\}\frac{1}{r^2}
\nonumber\\&&
+\frac{1}{1920}\,\nu\,\bigg\{
\left[-180(103+6\pi^2)+(33944+225\pi^2)\nu\right]\pp
\nonumber\\&&
+27\left[60(2\pi^2-3)-(2216+25\pi^2)\nu\right]\np^2
\bigg\}\frac{1}{r^3}
+\frac{1}{96}\,(926-63\pi^2)\nu\frac{1}{r^4}
\nonumber\\&&
+\omega\left[\pp-3\np^2\right]\frac{\nu^2}{r^3}.
\eea
The number $\omega$ in Eq.\ (\ref{r4reg}) is unknown due to the non-uniqueness
of the results of the regularization procedure, as discussed in the next
section.

The integrals given by $\widetilde{H}_{35}$ and $\widetilde{H}_{36}$ [defined
in Eqs.\ (\ref{r5def}) and (\ref{r6def})] are integrals of full divergences. To
discuss them let us first observe that the 3PN Hamiltonian given by Eqs.\
(\ref{r3sum})--(\ref{r6def}) not only describes the system of two point masses
but can be applied also to the system of two dusty bodies (provided the
following substitutions are made: $m_a\delta_a\to\rho_a$, $p_{ai}\delta_a\to
P_{ai}$, $({\bf p}_a^2/m_a)\delta_a\to{\bf P}^2_a/\rho_a$, etc., where $\rho_a$
and ${\bf P}_{a}$ are the mass and the linear momentum density of the $a$th
body, respectively). It is obvious that for dusty bodies the integrands in
$\widetilde{H}_{35}$ and $\widetilde{H}_{36}$ are locally integrable, so these
terms can contribute to the Hamiltonian only if the integrands fall off at
infinity slower than $1/r^4$.

Now we proceed as follows. We check whether the integrands in
$\widetilde{H}_{35}$ and $\widetilde{H}_{36}$ fall off at infinity at least as
$1/r^4$. If yes, the terms do not contribute to the Hamiltonian, if no, we must
study them in more detail. Because the asymptotic behaviour of all functions
entering $\widetilde{H}_{35}$ and $\widetilde{H}_{36}$ is the same for both
point masses and dusty bodies, we use the above prescription also for the
system of point masses, treating them as a limiting case of extended dusty
bodies.

One can check that the whole integrand in (\ref{r5def}) falls off at infinity
as $1/r^4$, so formally $\widetilde{H}_{35}$ does not contribute to the
Hamiltonian. For checking the consistency of our regularization procedures we
have integrated out all terms in $\widetilde{H}_{35}$ except the last two ones
(which contain the unknown functions ${\pitvii ij}$ and $\phi_{(8)2}$). To do
this we have used the regularization of Appendix B.2. We have obtained zeros
for all terms but the first one. This nonzero result is connected with the
ambiguity of the term $\widetilde{H}_{34}$, as explained in the next section.

The integrand in $\widetilde{H}_{36}$ does not fall off at infinity fast enough
not to possibly contribute to the Hamiltonian.  The reason is that the function
${\httvi ij}$ is a sum $h_{(6)ij}^{\rm TT div} + h_{(6)ij}^{\rm TT conv}$,
where $h_{(6)ij}^{\rm TT div}\sim r$ and $h_{(6)ij}^{\rm TT conv}\sim 1/r$ as
$r\to\infty$ (cf.\ Eqs.\ (\ref{htt6a}) and (\ref{htt6b})). The integrand in
$\widetilde{H}_{36}$ for ${\httvi ij}=h_{(6)ij}^{\rm TT conv}$ falls off at
infinity as $1/r^4$ but it falls off only as $1/r^2$ for ${\httvi
ij}=h_{(6)ij}^{\rm TT div}$. Therefore we have calculated $\widetilde{H}_{36}$
for ${\httvi ij}=h_{(6)ij}^{\rm TT div}$. We have used the regularization
procedures from Appendixes B.1 and B.2 (the need to use the Hadamard's
procedure of Appendix B.1 arises because $\Delta{\httiv ij}$ contains some
terms with Dirac deltas). To calculate $\Delta{\httiv ij}$ properly we also had
to employ the rule of differentiation of homogeneous functions from the
Appendix B.4. The result is zero.

On the basis of the above discussion we put the integrals given by
$\widetilde{H}_{35}$ and $\widetilde{H}_{36}$ equal to zero,
\be
\label{r56reg}
\widetilde{H}_{35} = \widetilde{H}_{36} = 0,
\ee
and adjust the regularized expression different from zero to the Hamiltonian
$\widetilde{H}_{34}$.

Collecting the Eqs.\ (\ref{r1reg})--(\ref{r56reg}) we finally obtain the autonomous
higher order 3PN Hamilton function $\widehat{H}_{\rm 3PN}$. It reads
\bea
\label{h3pnf}
\widehat{H}_{\rm 3PN}\left({\bf r},{\bf p},{\bf v},{\bf q}\right)
&=&\frac{1}{128}\left(-5+35\nu-70\nu^2+35\nu^3\right)\ppp^4
\nonumber\\&&\kern-9em
+\frac{1}{16}\bigg\{
\left(-7+42\nu-53\nu^2-6\nu^3\right)\ppp^3
+(1-2\nu)\nu^2\left[2\np^2\ppp^2+3\np^4\pp\right]
\bigg\}\frac{1}{r}
\nonumber\\&&\kern-9em
+\frac{1}{48}\bigg\{
3\left(-27+140\nu+96\nu^2\right)\ppp^2
+6(8+25\nu)\nu\np^2\pp-(35-267\nu)\nu\np^4
\bigg\}\frac{1}{r^2}
\nonumber\\&&\kern-9em
+\frac{1}{1536}\, \bigg\{
\left[-4800-3(8944-315\pi^2)\nu-7136\nu^2\right]\pp
+9\left(2672-315\pi^2+224\nu\right)\nu\np^2
\bigg\}\frac{1}{r^3}
\nonumber\\&&\kern-9em
+\frac{1}{96}\, \bigg\{12+(884-63\pi^2)\nu\bigg\}\frac{1}{r^4}
+\widehat{H}_{33}\left({\bf r},{\bf p},{\bf v},{\bf q}\right)
+\omega\left[\pp-3\np^2\right]\frac{\nu^2}{r^3},
\eea
where the number $\omega$ is unknown and the function
$\widehat{H}_{33}\left({\bf r},{\bf p},{\bf v},{\bf q}\right)$ is given by the
right-hand side of Eq.\ (\ref{r3reg}).

\section{Ambiguity}

The source of ambiguity given by the regularization procedure described in
Appendix B.2 can be explained as follows. Via integration by parts, some terms
in $\widetilde{H}_{34}$ can be represented in different ways. The regularization method applied to both
representations give different results.

As an example let us consider the integral of
$\phi_{(2)}\phi_{(2),i}\phi_{(2),j}{\httiv ij}$ which, by means of integration 
by parts, can be written as integral of 
$-\frac{1}{2}\phi_{(2)}^2\phi_{(2),ij}{\httiv ij}$ (in the Appendix B.4 another
such term is treated).  Application of the regularization procedure to the
difference between these  integrals (the integrand is a full divergence) gives
(in the reduced variables)
\be
\label{amb1}
\frac{1}{\mu}
\myinta\left(\frac{1}{2}\phi_{(2)}^2\phi_{(2),j}{\httiv ij}\right)_{,i}
=\frac{32}{5}\left[\pp-3\np^2\right]\frac{\nu^2}{r^3}.
\ee
Only if the result (\ref{amb1}) would be zero, application of the
regularization procedure to the integrals of
$\phi_{(2)}\phi_{(2),i}\phi_{(2),j}{\httiv ij}$ and
$-\frac{1}{2}\phi_{(2)}^2\phi_{(2),ij}{\httiv ij}$ would give the same results.

We can also obtain the result (\ref{amb1}) in a different way. Let us denote
the integrand from the left-hand side of (\ref{amb1}) by $F_{i,i}$ and let us
consider the volume integral
\be
\label{amb1a}
\int\limits_{B\left(0,R\right)\setminus 
\left[B\left({\bf x}_1,\ve_1\right)\cup B\left({\bf x}_2,\ve_2\right)\right]}
\kern-11ex d^3\!x\,F_{i,i},
\ee
where $B\left({\bf x}_a,\ve_a\right)$ ($a=1,2$) is a ball of radius $\ve_a$
around the position ${\bf x}_a$ of the $a$th body and $B\left(0,R\right)$
is a ball of radius $R$ centered at the origin of the coordinate system. We
apply Gauss's theorem to the integral (\ref{amb1a}) and then we calculate the
limit
\be
\label{amb1b}
\lim_{\ve_1\rightarrow0}
\kern-2ex \oint\limits_{\pa B\left({\bf x}_1,\ve_1\right)}
\kern-3ex d\sigma_i\,F_i
+ \lim_{\ve_2\rightarrow0}
\kern-2ex \oint\limits_{\pa B\left({\bf x}_2,\ve_2\right)}
\kern-3ex d\sigma_i\,F_i
+ \lim_{R\rightarrow\infty}
\kern-2ex \oint\limits_{\pa B\left(0,R\right)}
\kern-2ex d\sigma_i\,F_i
\ee
with the normal vectors pointing inwards the spheres $\pa B\left({\bf
x}_a,\ve_a\right)$ and outwards the sphere $\pa B\left(0,R\right)$. It is easy
to check that the integral over the sphere $\pa B\left(0,R\right)$ vanishes in
the limit $R\to\infty$ whereas the integrals over the spheres $\pa B\left({\bf
x}_a,\ve_a\right)$ diverge as $\ve_a\to0$, so to calculate them we use the
Hadamard's procedure from Appendix B.1. The result coincides with that given on
the right-hand side of Eq.\ (\ref{amb1}).

We have checked that for all these terms entering $\widetilde{H}_{34}$ for which
one can use different, via integration by parts, representations the ambiguity
is {\em  always} a multiple of the quantity
\be
\label{amb2}
\left[\pp - 3\np^2\right]\frac{\nu^2}{r^3}
= - (\nu p_i \pa_i)^2 \frac{1}{r},
\ee 
or, written in non-reduced form and with the momenta substituted through
the velocities (apart from a factor $4$),
\be
\label{amb3}
\frac{m_1v_1^iv_1^j}{2c^2} r^2_{s1} {\pa_{1i}}{\pa_{1j}}
\left(-\frac{Gm_2}{r_{12}}\right) +
\frac{m_2v_2^iv_2^j}{2c^2} r^2_{s2} {\pa_{2i}}{\pa_{2j}}
\left(-\frac{Gm_1}{r_{12}}\right),
\ee 
where $r_{s1}$ and $r_{s2}$ denote the Schwarschild radii ($r_{sa} =
Gm_a/2c^2$, in
isotropic coordinates) of the bodies 1 and
2, respectively. This expression indicates that the radius of the bodies might
come into play already at 3PN as indicated e.g.\ in Ref.\ \cite{S95}. The
interaction described by the terms (\ref{amb3}) is the interaction of the
(Newtonian) kinetic energy tensor of each body with the (Newtonian) tidal
potential of the other body scaled to the respective Schwarzschild radius.

We have tried to resolve the ambiguity (\ref{amb2}) in several ways. Firstly
we have extracted the local non-integrabilities in $\widetilde{H}_{34}$. Let
us denote by $F$ the total integrand of the $\widetilde{H}_{34}$ and by $
F_{sa}$ ($a=1,2$) the non-integrable part (i.e., of the order $1/r^3$ or
higher) of the Laurent expansion of the $F$ around the position ${\bf x}_a$ of
the $a$th point mass. Then we replace $F$ by the locally integrable
expression $F-\left(F_{s1}+F_{s2}\right)$. We have checked that the results of
applying the regularization procedure of Appendix B.2 are the same for $F$ and
$F-\left(F_{s1}+F_{s2}\right)$ (so the regularized value of the singular part
$F_{s1}+F_{s2}$ of the integrand is zero). This is so because by subtracting
the local non-integrabilities we have transferred the integrand, initially
locally non-integrable but integrable at infinity, to an integrand locally
integrable but non-integrable at infinity.

We have studied two further possibilities to overcome the ambiguity. The first
one, based on the Riesz's kernel representation of the Dirac delta
distribution, is described in detail in Appendix B.3. The Riesz's kernel
regularization of the divergence (\ref{amb1}) gives zero, i.e.\ it takes into
account also the contributions coming from the points ${\bf x}_a$. We were yet
not able to compute all terms in  $\widetilde{H}_{34}$ using the Riesz's kernel
regularization because of serious calculational problems, as described in
Appendix B.3 (only the first and the momenta dependent part of the fifth term
out of six terms in $\widetilde{H}_{34}$ we could calculate, but with
ill-defined ($1/\ve$ and ln$r_{12}$) multiplication factors for the terms of
type (\ref{amb2})). In the second case we have tried to employ the rule of
differentiation of homogeneous functions coming from the distribution theory.
This is described in Appendix B.4. Here we also haven't fully succeeded in
removing the ambiguity.

We conjecture that the ambiguity has its origin in the zero extension of the
bodies. We started with point-like bodies but the formalism reacted in such a
manner that the Schwarzschild radii of the bodies got introduced. They,
however, are far beyond the applicability of the post-Newtonian approximation
scheme and thus, the result turned out to be ambiguous. In the future we shall
investigate our conjecture in more detail.

We close this section with another observation. According to the applied
canonical formalism one is allowed to put ${\htt ij}$ and ${\pitt ij}$ equal to
zero at any freely chosen initial time. The energy of the initial state is
given by ${H}_{\le {\rm 3PN}}({\bf x}_a,{\bf p}_a,{\htt ij}=0,{\pitt ij}=0)$.
It exactly contains the first term of $\widetilde{H}_{34}$. But this term is
ill-defined even with respect to the Riesz's kernel regularization procedure
(Appendix B.3). This is a further indication that the binary point-mass model
is ill-defined at the 3PN level.

\section{Comparison with the known results}

The 3PN Hamiltonian derived in Section 4 we now compare with the results known
in the literature. We have found only two such results: the test body limit
and the static part of the full Hamiltonian. Let us stress that these
comparisons do not fix the ambiguity present in our paper.

The Hamiltonian (expressed in the reduced variables defined in the beginning of
Section 4) describing a test body orbiting around a Schwarzschild black hole
reads \cite{WS93} (we restored the explicit dependence on the speed of light
$c$)
\be
\label{tph1}
\widehat{H}^{\rm test}=
c^2\left[
\left(1-\frac{1}{2c^2 r}\right)\left(1+\frac{1}{2c^2 r}\right)^{-1}
\sqrt{1+\left(1+\frac{1}{2c^2 r}\right)^{-4}\frac{\pp}{c^2}}-1\right].
\ee
Expanding ({\ref{tph1}) with respect to $1/c$ and taking the $1/c^{6}$ contribution
we obtain the 3PN Hamiltonian for a test body:
\be
\label{tph2}
\widehat{H}^{\rm test}_{\rm 3PN}=
-\frac{5\ppp^4}{128}
-\frac{7\ppp^3}{16r}
-\frac{27\ppp^2}{16r^2}
-\frac{25\pp}{8r^3}
+\frac{1}{8r^4}.
\ee
To obtain the test-body limit of the 3PN Hamiltonian of our two-body point-mass
system we have to substitute $\nu=0$ into the right-hand side of Eq.\
(\ref{h3pnf}) (note that the ambiguous term and the term $\widehat{H}_{33}$
vanish for $\nu=0$). The result coincides with (\ref{tph2}).

The static part of the Hamiltonian is defined by means of the conditions
\be
\label{hs2}
{\bf p}_a=0,\quad{\pitt ij}=0,\quad{\htt ij}=0.
\ee
In the paper \cite{KT72} one can find the following expression for the static
part of the many-body point-mass 3PN Hamiltonian:
\bea
\label{hs1}
H_{\rm 3PN}^{\rm static}&=&\frac{1}{8}\frac{1}{(16\pi)^2}
\sum_a\sum_b\sum_c\sum_d\sum_e m_a m_b m_c m_d m_e
\nonumber\\[2ex]&&
\times\int\ldots\int d^3\!x_1\ldots d^3\!x_5\,
\delta\left({\bf x}_1-{\bf x}_a\right)
\delta\left({\bf x}_2-{\bf x}_b\right)
\delta\left({\bf x}_3-{\bf x}_c\right)
\delta\left({\bf x}_4-{\bf x}_d\right)
\delta\left({\bf x}_5-{\bf x}_e\right)
\nonumber\\[2ex]&&
\times
\left(\frac{2}{r_{12}r_{23}r_{34}r_{45}}+\frac{4}{r_{12}r_{23}r_{34}r_{35}}
+\frac{1}{r_{12}r_{23}r_{24}r_{25}}\right),
\eea
where $r_{mn}:=|{\bf x}_m-{\bf x}_n|$.
We have calculated $H_{\rm 3PN}^{\rm static}$ from (\ref{hs1}) for a two-body
system. For this we have used the Hadamard's regularization procedure of
Appendix B.1. The result is (in reduced variables)
\be
\label{hs3}
\widehat{H}_{\rm 3PN}^{\rm static}
=\left(\frac{1}{8}+\frac{\nu}{2}\right)\frac{1}{r^4}.
\ee
To calculate the static part of the 3PN Hamiltonian in our approach, we have
to implement the conditions (\ref{hs2}). After that the quantities
$\widetilde{H}_{32}$--$\widetilde{H}_{36}$ from Eqs.\
(\ref{r2def})--(\ref{r6def}) vanish. Only some part of the expression
$\widetilde{H}_{31}$ of Eq.\ (\ref{r1def}) survives. Applying for this part
the regularization procedure from Appendix B.1 one obtains the formula
(\ref{hs3}).

\vspace{3ex}
\noindent{\bf Acknowledgments}

\vspace{2ex}
\noindent The authors gratefully acknowledge discussions with Sergei Kopeikin,
Jan Pawlowski, and Prof.\ Hans Triebel. P.\ J.\ is thankful to Prof.\ Gernot
Neugebauer for invitations to the Friedrich-Schiller-University and kind
hospitality. This work was supported by the Max-Planck-Ge\-sell\-schaft (Research
Unit ``Theory of Gravitation"), by the Max-Planck-Gesellschaft Grant No.\
02160-361-TG74,  and by the KBN Grant No.\ 2 P303D 021 11 (P.\ J.).

\appendix\section{Some explicit solutions}

In this appendix we cite the known approximate solutions of the constraint
equations (\ref{ce1}) and (\ref{ce2}). The functions $\phi_{(2)}$,
$\phi_{(4)}$, $\phi_{(6)}$ (implicitly), ${\pitiii ij}$, and ${\httiv ij}$ can
be found e.g.\ in Refs.\ \cite{S85} and \cite{OOKH74}. The functions
$\phi_{(8)1}$, ${\pitv ij}$, and  $h_{(6)ij}^{\rm TT div}$ have been calculated
by us.

In most of the Poisson equations given below there are source terms of the form
$\sum_a f({\bf x})\delta_a$, where the function $f$ is singular at ${\bf
x}={\bf x}_a$. The Poisson integrals for these terms we have calculated as
follows
\be
\Delta^{-1}\left\{\sum_a f({\bf x})\delta_a\right\}
= \Delta^{-1}\left\{\sum_a f_{\rm reg}({\bf x}_a)\delta_a\right\}
= \sum_a f_{\rm reg}({\bf x}_a)\Delta^{-1}\delta_a
= -\frac{1}{4\pi}\sum_a f_{\rm reg}({\bf x}_a)\frac{1}{r_a},
\ee
where the regularized value $f_{\rm reg}$ of the function $f$ is defined in
Eq.\ (\ref{HPFdef}) of Appendix B.1. The Poisson integrals which do not contain
Dirac deltas we have computed using the explicit formulae
(\ref{il1})--(\ref{il6}) for inverse Laplacians listed at the end of this
appendix.

The Hamiltonian constraint equation (\ref{ce1}) is explicitly solved up to the
$1/c^6$ order. The solutions of Eqs.\ (\ref{lapphi2}) and (\ref{lapphi4}) for 
the functions $\phi_{(2)}$ and $\phi_{(4)}$ are known for $n$-body point-mass
system. They read
\bea
\phi_{(2)} &=& \frac{1}{4\pi}\sum_a\frac{m_a}{r_a},
\\[2ex]
\phi_{(4)} &=& -\frac{2}{(16\pi)^2}\sum_a\sum_{b\ne a}\frac{m_a m_b}{r_{ab}r_a}
+\frac{1}{8\pi}\sum_a\frac{{\bf p}_a^2}{m_a r_a}.
\eea

The solution of Eq.\ (\ref{lapphi6}) for the function $\phi_{(6)}$ is fully 
known only for two-body point-mass systems. It can be written as
\be
\phi_{(6)}=\phi_{(6)1}+\phi_{(6)2}+\phi_{(6)3},
\ee
where
\bea
\label{phi61}
\phi_{(6)1}&:=&\Delta^{-1}\left\{\sum_a\left(
-\frac{1}{64}\phi_{(2)}^2+\frac{1}{8}\phi_{(4)}
+\frac{5}{16}\phi_{(2)}\frac{{\bf p}_a^2}{m_a^2}
+\frac{1}{8}\frac{({\bf p}_a^2)^2}{m_a^4}\right)m_a\delta_a\right\},
\\[2ex]
\label{phi62}
\phi_{(6)2}&:=&-\Delta^{-1}\left\{\left({\pitiii ij}\right)^2\right\},
\\[2ex]
\label{phi63}
\phi_{(6)3}&:=&\frac{1}{2}\Delta^{-1}\left\{\phi_{(2),ij}{\httiv ij}\right\}
=\frac{1}{8\pi}\sum_a m_a 
\Delta^{-1}\left\{\left(\frac{1}{r_a}\right)_{,ij}{\httiv ij}\right\}.
\eea
Computation of the Poisson integrals from Eqs.\ (\ref{phi61}) and (\ref{phi62}) 
gives the results
\bea
\phi_{(6)1}&=&-\frac{1}{32\pi}\sum_a\frac{\pasqp^2}{m_a^3 r_a}
-\frac{1}{\left(16\pi\right)^2}\sum_a\sum_{b\ne a}\frac{m_b\pasq}{m_a r_{ab}}
\left(\frac{5}{r_a}+\frac{1}{r_b}\right)
\nonumber\\[2ex]&&
+\frac{1}{\left(16\pi\right)^3}\sum_a\sum_{b\ne a}\frac{m_a^2 m_b}{r_{ab}^2}
\left(\frac{1}{r_a}+\frac{1}{r_b}\right),
\\[2ex]
\phi_{(6)2}&=&-\frac{9}{8}\frac{1}{\left(16\pi\right)^2}
\sum_a\left(3\frac{\pasq}{r_a^2}+p_{ai}p_{aj}\left(\ln r_a\right)_{,ij}\right)
\nonumber\\[2ex]&&
+\frac{1}{\left(16\pi\right)^2}\sum_a\sum_{b\ne a}\left\{
-\frac{1}{4}\pana\pbnb\nanb^2\left(\Delta^{-1}r_a r_b\right)
\right.\nonumber\\[2ex]&&
+\left[-8\papb\nanb+3\pana\pbnb-8\panb\pbna\right]
\left(\Delta^{-1}\frac{1}{r_a r_b}\right)
\nonumber\\[2ex]&&\left.
+4\pana\pbna\nanb\left(\Delta^{-1}\frac{r_a}{r_b}\right)\right\}.
\eea
The Poisson integral needed to calculate $\phi_{(6)3}$ for two-body point-mass 
systems reads (here $b\ne a$)
\bea
\Delta^{-1}\left\{\left(\frac{1}{r_a}\right)_{,ij}{\httiv ij}\right\}
&=&\frac{1}{32\left(16\pi\right)^2}
\frac{m_a m_b}{r_a^3 r_b r_{ab}^3}\left(
3 r_a^4 - 12 r_a^3 r_{ab} + 18 r_a^2 r_{ab}^2 - 12 r_a r_{ab}^3 + 3 r_{ab}^4
\right.\nonumber\\[2ex]&&
+ 28 r_a^3 r_b
- 12 r_a^2 r_{ab} r_b - 12 r_a r_{ab}^2 r_b + 28 r_{ab}^3 r_b
- 30 r_a^2 r_b^2 + 60 r_a r_{ab} r_b^2
\nonumber\\[2ex]&&\left.
- 30 r_{ab}^2 r_b^2 - 36 r_a r_b^3
- 36 r_{ab} r_b^3 + 35 r_b^4 \right)
\nonumber\\[2ex]&&
+\frac{3}{64\pi}\frac{1}{m_a r_a^2}\left(\pasq-3\uapa^2\right)
+\frac{1}{32\pi}\frac{1}{m_b}\left\{
2\pbnb^2\frac{r_b}{r_a}
-4\frac{\pbsq}{r_a r_b}
\right.\nonumber\\[2ex]&&
+\left[2\pbsq\nnb^2
-4\pbn\pbnb\nnb
\right.\nonumber\\[2ex]&&\left.
-4\pbn\pbnb\nnb\right]
\left(\Delta^{-1}\frac{r_b}{r_a}\right)
+8\pbn^2\left(\Delta^{-1}\frac{1}{r_a r_b}\right)
\nonumber\\[2ex]&&\left.
+\frac{1}{6}\pbnb^2\nnb^2\left(\Delta^{-1}\frac{r_b^3}{r_a}\right)\right\}.
\eea

Only part of the solution of Eq.\ (\ref{lapphi8}) for the function $\phi_{(8)}$ 
is known, namely
\be
\label{phi81def}
\phi_{(8)1}
=\Delta^{-1}S_{(8)1}
+\frac{1}{2}\Delta^{-1}\left\{\phi_{(4),ij}{\httiv ij}\right\}
+\frac{1}{2}\left({\httiv ij}\right)^2,
\ee
where
\bea
S_{(8)1}&:=&\sum_a\left\{
\frac{1}{512}\phi_{(2)}^3-\frac{1}{32}\phi_{(2)}\phi_{(4)}
+\frac{1}{8}\phi_{(6)}
+\left(\frac{5}{16}\phi_{(4)}-\frac{15}{128}\phi_{(2)}^2\right)
\frac{{\bf p}_a^2}{m_a^2}
\right.\nonumber\\[2ex]&&\left.
-\frac{9}{64}\phi_{(2)}\frac{({\bf p}_a^2)^2}{m_a^4}
-\frac{1}{16}\frac{({\bf p}_a^2)^3}{m_a^6}
+\frac{1}{2}\frac{p_{ai}p_{aj}}{m_a^2}{\httiv ij}\right\}m_a\delta_a.
\eea
We have computed the Poisson integrals from Eq.\ (\ref{phi81def}). They are
given by
\bea
\Delta^{-1}S_{(8)1}
&=&\frac{1}{64\pi}\sum_a\frac{\pasqp^3}{m_a^5 r_a}
+\frac{1}{4\left(16\pi\right)^2}\sum_a\sum_{b\ne a}\left\{
\frac{m_a\pasqp^2}{m_b^3 r_{ab}}\left(\frac{1}{r_a}+\frac{9}{r_b}\right)\right\}
\nonumber\\[2ex]&&
-\frac{1}{2\left(16\pi\right)^2}\sum_a\sum_{b\ne a}
\frac{8\pasqp^2+2\uabpa^2\pasq+3\uabpa^4}{m_a m_b r_{ab}r_a}
\nonumber\\[2ex]&&
+\frac{1}{8\left(16\pi\right)^3}\sum_a\sum_{b\ne a}\frac{m_a}{r_{ab}^2}
\left\{\pasq\left(\frac{47}{r_a}-\frac{6}{r_b}\right)
+15\uabpa^2\left(\frac{1}{r_a}+\frac{6}{r_b}\right)\right\}
\nonumber\\[2ex]&&
+\frac{5}{2\left(16\pi\right)^3}\sum_a\sum_{b\ne a}
\frac{m_a^2\pasq}{m_b r_{ab}^2}\left(\frac{1}{r_a}+\frac{3}{r_b}\right)
\nonumber\\[2ex]&&
-\frac{1}{2\left(16\pi\right)^4}\sum_a\sum_{b\ne a}
\left\{\frac{5m_a^2 m_b^2}{r_{ab}^3 r_a}
+\frac{m_a^3 m_b}{r_{ab}^3}\left(\frac{1}{r_a}+\frac{1}{r_b}\right)\right\},
\nonumber\\[2ex]
\frac{1}{2}\Delta^{-1}\left\{\phi_{(4),ij}{\httiv ij}\right\}
&=&\frac{1}{16\pi}\sum_a\frac{\pasq}{m_a}
\Delta^{-1}\left\{\left(\frac{1}{r_a}\right)_{,ij}{\httiv ij}\right\}
\nonumber\\[2ex]&&
-\frac{1}{\left(16\pi\right)^2}\sum_a\sum_{b\ne a}
\frac{m_a m_b}{r_{ab}}
\Delta^{-1}\left\{\left(\frac{1}{r_a}\right)_{,ij}{\httiv ij}\right\}.
\eea

The momentum constraint equations (\ref{eqpi3})--(\ref{eqpi71}) can be written
in the form
\be
\label{piti1}
{{\piti ij}}_{,j} = S^i.
\ee
Making use of the decomposition ({\ref{pitidec}) it is not difficult to write
the solution of Eq.\ (\ref{piti1}) in the form
\be
\label{piti2}
{\piti ij} = -\frac{1}{2}\Theta^{ij}_k S^k,
\ee
where the operator $\Theta^{ij}_k$ is defined as follows
\be
\Theta^{ij}_k :=
\left(
- \frac{1}{2} \delta_{ij} \pa_k 
+ \delta_{ik} \pa_j
+ \delta_{jk} \pa_i 
- \frac{1}{2} \pa_i \pa_j \pa_k \Delta^{-1}
\right) \Delta^{-1}.
\ee

Using Eq.\ (\ref{piti2}) we have obtained the solutions of Eqs.\ (\ref{eqpi3})
and (\ref{eqpi5}) valid for $n$-body point-mass systems. The solution of
Eq.\ (\ref{eqpi3}) reads
\be
{\pitiii ij}=\frac{1}{16\pi}\sum_a p_{ak}\left\{
-\delta_{ij}\left(\frac{1}{r_a}\right)_{\kern-1.5mm,k}
+2\left[\delta_{ik}\left(\frac{1}{r_a}\right)_{\kern-1.5mm,j}
+\delta_{jk}\left(\frac{1}{r_a}\right)_{\kern-1.5mm,i}\right]
-\frac{1}{2}r_{a,ijk}
\right\}.
\ee
The function $\pi_{(3)}^i$, connected with the function ${\pitiii ij}$ by means 
of Eq.\ (\ref{pitidec}), is equal to
\be
\pi_{(3)}^i=\frac{1}{8\pi}\sum_a p_{ai}\frac{1}{r_a}
-\frac{1}{32\pi}\sum_a p_{aj}r_{a,ij}.
\ee
The explicit formula for the function ${\pitv ij}$ reads
\bea
{\pitv ij}&=&\frac{3}{\left(16\pi\right)^2}\sum_a\frac{m_a}{r_a^3}\left[
\left(p_{ai}n_a^j+p_{aj}n_a^i\right)-\uapa\delta_{ij}+\uapa n_a^i n_a^j\right]
\nonumber\\[2ex]&&
+\frac{1}{\left(16\pi\right)^2}\sum_a\sum_{b\ne a}m_b\left\{
-\delta_{ij}p_{ak}\left(\frac{1}{r_a}\right)_{,k}\frac{1}{r_b}
+\left[-4\delta_{ij}\pan\nna-3\pa_i\pa_j\pana
\right.\right.\nonumber\\[2ex]&&\left.
+4\pan\left(\pa_i\pa_{aj}+\pa_j\pa_{ai}\right)
+4\left(p_{aj}\pa_i+p_{ai}\pa_j\right)\nna\right]
\left(\Delta^{-1}\frac{1}{r_a r_b}\right)
\nonumber\\[2ex]&&
+\left[\frac{1}{2}\delta_{ij}\pana\nna^2
-\left(\pa_i\pa_{aj}+\pa_j\pa_{ai}\right)\pana\nna\right]
\left(\Delta^{-1}\frac{r_a}{r_b}\right)
\nonumber\\[2ex]&&\left.
-4\pa_i\pa_j\pan\nna\left(\Delta^{-2}\frac{1}{r_a r_b}\right)
+\frac{1}{2}\pa_i\pa_j\pana\nna^2\left(\Delta^{-2}\frac{r_a}{r_b}\right)
\right\}.
\eea

The TT-part of the metric in the leading order ($1/c^4$) is the solution of
Eq.\ (\ref{httiv}). It is given by (here $s_{ab}:=r_{a}+r_{b}+r_{ab}$; be aware that
in Eq.\ (20) in \cite{JS97} there is a misprint)
\bea
\label{h4TT}
h^{\rm TT}_{(4)ij}&=&
\frac{1}{4}\frac{1}{16\pi}
\sum\limits_a \frac{1}{m_a r_a} \left\{
\Big[ {\bf p}_a^2 - 5 ({\bf n}_{a}\cdot{\bf p}_a)^2 \Big] \delta_{ij}
+ 2 p_{ai} p_{aj}
\right.\nonumber\\[2ex]&&\left.
+ \Big[3 ({\bf n}_{a}\cdot{\bf p}_a)^2 - 5{\bf p}_a^2 \Big] n_{a}^i n_{a}^j
+ 6 ({\bf n}_{a}\cdot{\bf p}_a) ( n_a^i p_{aj} + n_a^j p_{ai} ) \right\}
\nonumber\\[2ex]&&
+ \frac{1}{8}\frac{1}{\left(16\pi\right)^2}
\sum\limits_a\sum\limits_{b\ne a} m_a m_b
\left\{ - \frac{32}{s_{ab}}
\left(\frac{1}{r_{ab}} + \frac{1}{s_{ab}}\right) n_{ab}^i n_{ab}^j
\right.\nonumber\\[2ex]&&
+ 2 \left(\frac{r_a + r_b}{r_{ab}^3} + \frac{12}{s_{ab}^2}\right) n_a^i n_{b}^j
+ 16 \left( \frac{2}{s_{ab}^2} - \frac{1}{r_{ab}^2} \right)
\left( n_a^i n_{ab}^j +  n_a^j n_{ab}^i \right)
\nonumber\\[2ex]&&
+ \left[ \frac{5}{r_{ab}r_a}
- \frac{1}{r_{ab}^3}\left(\frac{r_b^2}{r_a}+3r_a\right)
- \frac{8}{s_{ab}}
\left(\frac{1}{r_a} + \frac{1}{s_{ab}}\right) \right] n_a^i n_{a}^j
\nonumber\\[2ex]&&\left.
+ \left[ 5 \frac{r_a}{r_{ab}^3} \left( \frac{r_a}{r_b} - 1 \right)
- \frac{17}{r_{ab}r_a} + \frac{4}{r_a r_b}
+ \frac{8}{s_{ab}} \left( \frac{1}{r_a} + \frac{4}{r_{ab}} \right) \right]
\delta_{ij} \right\}.
\eea

The part of the TT-metric at $1/c^6$ which diverges linearly at infinity is
given by Eq.\ (\ref{htt6b}). After computing the double Poisson integral in
(\ref{htt6b}) one finds
\bea
\label{h6TT}
h^{\rm TT div}_{(6)ij}&=&
\frac{1}{24}\frac{1}{16\pi}\frac{\pa^2}{\pa t^2}
\sum\limits_a \frac{r_a}{m_a} \left\{
\Big[7({\bf n}_{a}\cdot{\bf p}_a)^2-11{\bf p}_a^2\Big] \delta_{ij}
+ 26 p_{ai} p_{aj}
\right.\nonumber\\[2ex]&&\left.
+ \Big[7{\bf p}_a^2-({\bf n}_{a}\cdot{\bf p}_a)^2\Big] n_{a}^i n_{a}^j
- 10({\bf n}_{a}\cdot{\bf p}_a) ( n_a^i p_{aj} + n_a^j p_{ai} ) \right\}
\nonumber\\[2ex]&&
- \frac{1}{\left(16\pi\right)^2}\frac{\pa^2}{\pa t^2}
\sum\limits_a\sum\limits_{b\ne a} m_a m_b\left\{
\frac{1}{8}\left[ \frac{1}{r_{ab}^3}\left(r_a^3-r_a^2 r_b\right)
+ \frac{5r_a}{r_{ab}} \right]\delta_{ij}
\right.\nonumber\\[2ex]&&
+\frac{1}{12}\left(
\pa_i\pa_{bj}+\pa_j\pa_{bi}-\pa_i\pa_{aj}-\pa_j\pa_{ai}
\right)\frac{r_a^3}{r_{ab}}
+\frac{1}{144r_{ab}^3}\pa_i\pa_j\left[r_a^5-r_a^3(r_b^2+17r_{ab}^2)\right]
\nonumber\\[2ex]&&\left.
- \frac{1}{2}\delta_{ij}\left(\Delta^{-1}\frac{1}{r_a r_b}\right)
+ \frac{1}{2}\left(8\pa_{ai}\pa_{bj}-\pa_i\pa_j\right)
\left(\Delta^{-2}\frac{1}{r_a r_b}\right)
\right\}.
\eea

Below we list explicit formulae for the inverse Laplacians which have been used
throughout this appendix:
\bea
\label{il1}
\Delta^{-1}\frac{1}{r_a r_b}&=&\ln s_{ab},
\\[2ex]
\Delta^{-2}\frac{1}{r_a r_b}&=& 
\frac{1}{36}\left(-r_a^2 + 3 r_a r_{ab} + r_{ab}^2 - 3 r_a r_b
+ 3 r_{ab} r_b - r_b^2 \right)
+\frac{1}{12} \left( r_a^2 - r_{ab}^2 + r_b^2 \right)\ln s_{ab},
\\[2ex]
\Delta^{-1}\frac{r_a}{r_b}&=&
\frac{1}{18}\left(
-r_a^2 - 3r_ar_{ab} - r_{ab}^2 + 3r_ar_b + 3r_{ab}r_b + r_b^2\right)
+\frac{1}{6}\left(r_a^2 + r_{ab}^2 - r_b^2\right)\ln s_{ab},
\\[2ex]
\Delta^{-2}\frac{r_a}{r_b}&=&
\frac{1}{7200}\left(
-r_a^4 - 30r_a^3r_{ab} - 62r_a^2r_{ab}^2 + 90r_ar_{ab}^3 + 63r_{ab}^4 - 
30r_a^3r_b
+ 30r_a^2r_{ab}r_b
\right.\nonumber\\[2ex]&&
- 90r_ar_{ab}^2r_b
+ 90r_{ab}^3r_b - 
62r_a^2r_b^2 - 90r_ar_{ab}r_b^2 - 126r_{ab}^2r_b^2 + 90r_ar_b^3
\nonumber\\[2ex]&&\left.
+90r_{ab} r_b^3+63r_b^4\right)
+\frac{1}{240}\left(r_a^4 + 2r_a^2r_{ab}^2 - 3r_{ab}^4 +
2r_a^2r_b^2 + 6r_{ab}^2r_b^2 - 3r_b^4\right)\ln s_{ab},
\\[2ex]
\Delta^{-1}r_a r_b&=&
\frac{1}{3600}\left(63r_a^4 + 90r_a^3r_{ab} - 62r_a^2r_{ab}^2 - 30r_ar_{ab}^3 -
r_{ab}^4 +  90r_a^3r_b - 90r_a^2r_{ab}r_b+ 30r_ar_{ab}^2r_b
\right.\nonumber\\[2ex]&&\left.
- 30r_{ab}^3r_b - 
126r_a^2r_b^2 - 90r_ar_{ab}r_b^2 - 62r_{ab}^2r_b^2 + 90r_ar_b^3 + 
90r_{ab}r_b^3 + 63r_b^4\right)
\nonumber\\[2ex]&&
+\frac{1}{120}\left(-3r_a^4 + 2r_a^2r_{ab}^2 +
r_{ab}^4 + 6r_a^2r_b^2 + 2r_{ab}^2r_b^2 - 3r_b^4\right)\ln s_{ab},
\\[2ex]
\label{il6}
\Delta^{-1}\frac{r_a^3}{r_b}&=&
\frac{1}{1200}\left(-63r_a^4 - 90r_a^3r_{ab} - 2r_a^2r_{ab}^2 - 90r_ar_{ab}^3 -
63r_{ab}^4 +  150r_a^3r_b
\right.\nonumber\\[2ex]&&
+ 90r_a^2r_{ab}r_b+ 90r_ar_{ab}^2r_b
+150r_{ab}^3r_b +  126r_a^2r_b^2 + 90r_ar_{ab}r_b^2
\nonumber\\[2ex]&&\left.
+ 126r_{ab}^2r_b^2
-90r_ar_b^3 -  90r_{ab}r_b^3 - 63r_b^4\right)
\nonumber\\[2ex]&&
+\frac{1}{40}\left(3r_a^4+2r_a^2r_{ab}^2+3r_{ab}^4-6r_a^2 r_b^2-6r_{ab}^2 r_b^2
+3r_b^4\right)\ln s_{ab}.
\eea

\section{Regularization procedures}

In this appendix we describe techniques which can be used to regularize
integrals which appear in our paper. More details are given in Ref.\
\cite{J97}.

\subsection{Hadamard's ``partie finie" regularization}

Let $f$ be a real valued function defined in a neighbourhood of a point
${\bf x}_o\in\mbox{\af R}^3$, excluding this point. At ${\bf x}_o$ the
function $f$ is assumed to be singular. We define the family of auxiliary
complex valued functions $f_{\bf n}$ (labelled by unit vectors ${\bf n}$) in
the following way:
\be
f_{\bf n}: \mbox{\af C}\ni\ve\mapsto f_{\bf n}(\ve) :=
f\left({\bf x}_o+\ve{\bf n}\right) \in \mbox{\af C}.
\ee
We expand $f_{\bf n}$ into a Laurent series around $\ve = 0$:
\be
\label{laurent}
f_{\bf n}(\ve) = \sum\limits_{m = -N}^{\infty}a_m({\bf n})\,\ve^m.
\ee
The coefficients $a_m$ in this expansion depend on the unit vector ${\bf n}$. We
define the regularized value of the function $f$ at ${\bf x}_o$ as the
coefficient of $\ve^0$ in the expansion (\ref{laurent}) averaged over all
directions:
\be
\label{HPFdef}
f_{\rm reg}\left({\bf x}_o\right)
:=\frac{1}{4\pi}\oint\!d\Omega\,a_0({\bf n}).
\ee
We use formula (\ref{HPFdef}) to compute all integrals which contain Dirac
delta distribution. It means that we define
\be
\label{DDdef}
\myinta f\left({\bf x}\right)
\delta\left({\bf x}- {\bf x}_a\right) := f_{\rm reg}\left({\bf x}_a\right).
\ee

The procedure described here was used by one of the authors (cf.\ Appendix B
in \cite{S85}) in the calculation of the 2PN and 2.5PN ADM Hamiltonians for
many-body point-mass systems. More details on the applications of the
Hadamard's regularization can be found in \cite{J97}. Also the relation of the
Hadamard's procedure to the regularization described in Appendix B.2 and the
rule of differentiation of homogeneous functions of Appendix B.4 is discussed
in \cite{J97}.

\subsection{Riesz's formula based regularization}

The following formula, firstly derived by Riesz (see Eqs.\ (7) and (10) in
Chapter 2 of \cite{R49}), can serve as a tool to regularize a class of
divergent integrals of $r_1^\alpha\,r_2^\beta$ using the analytic continuation
arguments:
\be
\label{r1}
\left[\myinta r_1^\alpha\,r_2^\beta\right]_{\rm reg}:=
\pi^{3/2}
\frac{\mit\Gamma\left(\frac{\alpha+3}{2}\right)
\mit\Gamma\left(\frac{\beta+3}{2}\right)
\mit\Gamma\left(-\frac{\alpha+\beta+3}{2}\right)}
{\mit\Gamma\left(-\frac{\alpha}{2}\right)
\mit\Gamma\left(-\frac{\beta}{2}\right)
\mit\Gamma\left(\frac{\alpha+\beta+6}{2}\right)}\,r_{12}^{\alpha+\beta+3}.
\ee

In deriving the 3PN Hamiltonian for a two-body point-mass system more general
integrals than those which can be regularized by means of the formula
(\ref{r1}) appear. We succeeded in deriving the generalization of the formula
(\ref{r1}) for those integrals. The generalized formula reads
\be
\label{r2a}
\left[\myinta
r_1^\alpha\,r_2^\beta\left(r_1+r_2+r_{12}\right)^\gamma
\right]_{\rm reg}:=
R(\alpha,\beta,\gamma)\,r_{12}^{\alpha+\beta+\gamma+3},
\ee
where
\bea
\label{r2b}
R(\alpha,\beta,\gamma)&:=&2\pi
\frac{\mit\Gamma\left(\alpha+2\right)\mit\Gamma\left(\beta+2\right)
\mit\Gamma\left(-\alpha-\beta-\gamma-4\right)}{\mit\Gamma\left(-\gamma\right)}
\nonumber\\[2ex]&&
\times\left[
I_{1/2}\left(\alpha+2,-\alpha-\gamma-2\right)
+I_{1/2}\left(\beta+2,-\beta-\gamma-2\right)
\right.\nonumber\\[2ex]&&\left.
-I_{1/2}\left(\alpha+\beta+4,-\alpha-\beta-\gamma-4\right)-1
\right].
\eea
The function $I_{1/2}$ in Eq.\ (\ref{r2b}) is defined as follows
\be
I_{1/2}\left(x,y\right):=\frac{B_{1/2}\left(x,y\right)}{B\left(x,y\right)},
\ee 
where $B$ stands for the beta function (Euler's integral of the first kind) and
$B_{1/2}$ is the incomplete beta function; it can be expressed in terms of the
Gauss hypergeometric function $_2F_1$:
\be
B_{1/2}\left(x,y\right)=\frac{1}{2^x x}\,
{_2F_1}\!\!\left(1-y,x;x+1;\frac{1}{2}\right).
\ee

The regularization procedure based on the formulae (\ref{r2a}) and (\ref{r2b})
consists of several steps. We enumerate them now.
\begin{enumerate}
\item By means of Eqs.\ (\ref{cm}) and (\ref{v1v2}) we replace ${\bf p}_2$ by 
$-{\bf p}_1$, $\dot{\bf p}_2$ by $-\dot{\bf p}_1$, and $\dot{\bf x}_1$ and 
$\dot{\bf x}_2$ by ${\bf v}$.
\item We eliminate the unit vector ${\bf n}_2$ by the relation
\be
{\bf n}_2=\frac{r_1}{r_2}{\bf n}_1+\frac{r_{12}}{r_2}{\bf n}_{12}.
\ee
Then we expand the scalar product ${\bf n}_1\cdot{\bf n}_{12}$:
\be
{\bf n}_1\cdot{\bf n}_{12}=\frac{r_2^2-r_1^2-r_{12}^2}{2r_1 r_{12}}.
\ee
\item All integrands we reduce to integrands which depend
on $r_1$ and $r_2$ only. To do this we use a set of substitutions. For the
integrands of the type $\left({\bf n}_1\cdot{\bf p}_1\right)^k
f\left(r_1,r_2\right)$, $k=1,\ldots,6$, we use the rules (obtained by 
considering the integration in prolate spheroidal coordinates)
\bea
\left({\bf n}_1\cdot{\bf p}_1\right) f\left(r_1,r_2\right)&\to&
a f\left(r_1,r_2\right),
\\[2ex]
\left({\bf n}_1\cdot{\bf p}_1\right)^2 f\left(r_1,r_2\right)&\to&
\left(a^2+\frac{1}{2}b^2\right)f\left(r_1,r_2\right),
\\[2ex]
\left({\bf n}_1\cdot{\bf p}_1\right)^3 f\left(r_1,r_2\right)&\to&
\left(a^3+\frac{3}{2}a b^2\right)f\left(r_1,r_2\right),
\\[2ex]
\left({\bf n}_1\cdot{\bf p}_1\right)^4 f\left(r_1,r_2\right)&\to&
\left(a^4+3a^2 b^2+\frac{3}{8}b^4\right)f\left(r_1,r_2\right),
\\[2ex]
\left({\bf n}_1\cdot{\bf p}_1\right)^5 f\left(r_1,r_2\right)&\to&
\left(a^5+5a^3 b^2+\frac{15}{8}a b^4\right)f\left(r_1,r_2\right),
\\[2ex]
\left({\bf n}_1\cdot{\bf p}_1\right)^6 f\left(r_1,r_2\right)&\to&
\left(a^6+\frac{15}{2}a^4 b^2+\frac{45}{8}a^2 b^4+\frac{5}{16}b^6\right)
f\left(r_1,r_2\right),
\eea 
where
\bea
a&:=&\left({\bf n}_{12}\cdot{\bf p}_1\right)
\frac{r_2^2-r_1^2-r_{12}^2}{2r_1 r_{12}},
\\[2ex]
b&:=&\sqrt{{\bf p}_1^2-\left({\bf n}_{12}\cdot{\bf p}_1\right)^2}
\frac{\sqrt{\left[\left(r_1+r_2\right)^2-r_{12}^2\right]
\left[r_{12}^2-\left(r_1-r_2\right)^2\right]}}{2r_1 r_{12}}.
\eea
After this step the integrand becomes the linear combination of the type
\be
\label{r01}
\sum_{I=1}^{N} c_I r_1^{\alpha_I}r_2^{\beta_I}
\left(r_1+r_2+r_{12}\right)^{\gamma_I},
\ee
where $\alpha_I$, $\beta_I$, and $\gamma_I$ are integers; the constant
coefficients $c_I$ may depend on $r_{12}$, ${\bf p}_1^2$, 
$\dot{\bf p}_1^2$, ${\bf v}^2$,
$\left({\bf n}_{12}\cdot{\bf p}_1\right)$,
$\left({\bf n}_{12}\cdot\dot{\bf p}_1\right)$,
$\left({\bf n}_{12}\cdot{\bf v}\right)$,
$\left({\bf v}\cdot{\bf p}_1\right)$,
$\left({\bf v}\cdot\dot{\bf p}_1\right)$, and
$\left({\bf p}_1\cdot\dot{\bf p}_1\right)$.
\item We define the auxiliary function
\be
\label{r02}
J^{\mu,\nu}_{\ve}(\alpha,\beta,\gamma):=R(\alpha+\mu\ve,\beta+\nu\ve,\gamma)\,
r_{ab}^{\alpha+\beta+\gamma+(\mu+\nu)\ve+3}
\ee
and for the combination (\ref{r01}) we calculate the limit
\be
\label{r03}
\lim_{\ve\rightarrow0}\sum_{I=1}^N
c_I\,J^{\mu,\nu}_{\ve}(\alpha_I,\beta_I,\gamma_I).
\ee
It turns out that for all terms considered in the paper
this limit is of the form
\be
\label{r031}
A+B\,\frac{\mu}{\nu}+C\,\frac{\nu}{\mu},
\ee
where $A$, $B$, $C$ do not depend on $\mu$ and $\nu$.
\item As the regularized value of the integral of the sum (\ref{r01}) we
take the number $A$, i.e.\
\be
\label{r04}
\left[\myinta
\sum_{I=1}^{N} c_I r_1^{\alpha_I}r_2^{\beta_I}
\left(r_1+r_2+r_{12}\right)^{\gamma_I}
\right]_{\rm reg}:=A.
\ee
\end{enumerate}

\subsection{Riesz's kernel based regularization}

Three-dimensional Dirac delta distribution can be represented as the limit
\be
\label{rk1}
\delta({\bf x}-{\bf x}_a)
= \lim\limits_{\alpha \to 0} I^\alpha ( {\bf x}, {\bf x}_a ),
\ee
where $I^\alpha ( {\bf x}, {\bf x}' )$ is the Riesz's kernel defined as follows:
\be
\label{rk2}
I^\alpha ( {\bf x}, {\bf x}' ) :=
\frac{\ds \mit\Gamma \left( \frac{3-\alpha}{2} \right) }
     {\ds \pi^{3/2} 2^\alpha \mit\Gamma \left( \frac{\alpha}{2} \right) }
|{\bf x}-{\bf x}'|^{\alpha - 3}.
\ee
The regularization of a class of integrals based on the kernel (\ref{rk2}) and
its relation to Hadamard's procedure of Appendix B.1 is discussed by Riesz
\cite{R49}. For us, the Riesz's kernel regularization consists in using the
kernel (\ref{rk2}) instead of the Dirac delta in the source terms of the
constraint equations, solving these equations, and performing the
regularization of integrals needed to obtain the 3PN dynamics in the way
described below.

For a point-mass two-body system we need two families of Riesz's kernels, each
associated with the different mass. Let the first family be labelled by the
parameter $\alpha$, and the second one by $\beta$. After replacing the Dirac
deltas by  Riesz's kernels the right-hand sides of the constraint equations
(\ref{ce1})--(\ref{ce2}) depend on $\alpha$ and $\beta$. The first step of the 
procedure is to solve constraint equations iteratively keeping $\alpha$ and 
$\beta$ as unspecified complex parameters.

This is a very difficult task. Out of all equations for the functions 
$\phi_{(n)}$ we have fully solved only the simplest one for the function
$\phi_{(2)}$. Also only the leading order momentum constraint equation for
${\pitiii ij}$ we have  solved. The TT-part of the metric is not known even in
the lowest order --- only the  part depending quadratically on the momenta has
been calculated.

The solutions obtained by us are enough to calculate that part of the full
divergence (\ref{amb1}) (discussed in Section 5) which depends quadratically on
the momenta. The regularization of the  full divergence by means of the
procedure described in Appendix B.2 gives a non-zero  result.  If one uses the
Riesz's kernel generalization of the function $\phi_{(2)}$ and of the momenta
dependent part of the function ${\httiv ij}$ the result of regularization
of the full divergence using the procedure described below is zero.

The procedure is similar to that described in Appendix B.2. The first three
steps  of the regularization procedure from Appendix B.2 apply here. After that
all  integrands to be regularized are functions of $r_1$, $r_2$ and the
parameters $\alpha$ and $\beta$. So any integrand can be written as
\be
\label{rk3}
f\left(r_1,r_2,\alpha,\beta\right).
\ee
We need now to use formulae analogous to the generalized Riesz's formula
(\ref{r1})  from Appendix B.2 to perform integration of (\ref{rk3}) for any
$\alpha$ and  $\beta$. Let's denote the result of integration by
$R\left(\alpha,\beta\right)$.  Then the following limit is calculated:
\be
\label{rk4}
\lim_{\ve\rightarrow0}R(\alpha+\mu\ve,\beta+\nu\ve).
\ee
For the divergence (\ref{amb1}), to perform the integration it is enough to use 
the formula (\ref{r1}) from Appendix B.2 for $\gamma=0$. The limit 
(\ref{rk4}) comes out to be zero.

For all terms for which we were able to apply the above procedure we obtained
the limits (\ref{rk4}) of the form given in Eq.\ (\ref{r031}). However, we
were not able to check the cancellation of all divergent terms in
$\widetilde{H}_{34}$ after regularization because we were not able to
calculate all terms using the Riesz's kernel procedure. But again, the
ill-defined terms we have obtained were of the type (\ref{amb2}).

\subsection{The rule to differentiate homogeneous functions.}

In distribution theory there exists a rule to differentiate some homogeneous
and locally non-integrable functions under the integral sign \cite{GS64}. We
have studied how useful this rule is in our calculations.

Let $f$ be a real valued function defined in a neighbourhood of the origin in
{\af R}$^3$.  $f$ is said to be a positively homogeneous function of degree
$\lambda$, if for any number $a>0$
\be
f\left(a\,{\bf x}\right)=a^\lambda\,f\left({\bf x}\right).
\ee
Let $k:=-\lambda-2$. If $\lambda$ is an integer and if $\lambda\le-2$ (i.e.\
$k$ is a non-negative integer), then the partial derivative of $f$ with
respect to the coordinate $x^i$ has to be calculated by means of the formula
(cf.\ Eq.\ (5.15) in \cite{K85})
\be
\label{erd1}
\pa_i f =\left(\pa_i f\right)_{\rm ord}
+\frac{(-1)^k}{k!}\frac{\pa^k\delta({\bf x})}{\pa x^{i_1}\ldots\pa x^{i_k}}
\oint_\Sigma d\sigma_i\,f\,x^{i_1}\ldots x^{i_k},
\ee
where $\left(\pa_i f\right)_{\rm ord}$ means the derivative computed using
the standard rules of differentiations, $\Sigma$ is any smooth close surface
surrounding the origin and $d\sigma_i$ is the surface element on $\Sigma$.

As an example of applying the rule (\ref{erd1}) let us consider the full divergence
(connected with the fourth term in the $\widetilde{H}_{34}$ part of the 3PN
Hamiltonian, see Eq.\ (\ref{r4def}))
\bea
\label{erd2}
\myinta \left\{
\left(2\pi_{(3)}^i+\Delta^{-1}\pi_{(3),il}^l\right)
\left(2\pi_{(3)}^k+\Delta^{-1}\pi_{(3),km}^m\right)_{,j}{\httiv ij}
\right\}_{,k}.
\eea
Applying the regularization procedure from Appendix B.2 to the divergence
(\ref{erd2}) gives a result much more complicated than that given by Eq.\
(\ref{amb2}). After performing in (\ref{erd2}) differentiation with respect to
$x^k$ one finds that only in one term the rule (\ref{erd1}) can be applied.
The term reads
\be
\label{erd3}
3 \left(2\pi_{(3)}^i+\Delta^{-1}\pi_{(3),il}^l\right) {\httiv ij}
\pi_{(3),jk}^k.
\ee
The rule (\ref{erd1}) applied to $\pi_{(3),jk}^k$ yields
\be
\label{erd4}
\myinta \left\{
-\frac{1}{4} \left(2\pi_{(3)}^i+\Delta^{-1}\pi_{(3),il}^l\right) {\httiv ij}
\right\} \sum_a p_{aj} \delta_a.
\ee
The integral (\ref{erd4}) is calculated by means of Hadamard's procedure
from Appendix B.1. After adding the result of regularization of the integral
(\ref{erd4}) to the result of regularization of the integral (\ref{erd2}) we
obtain a multiple of the quantity (\ref{amb2}).

Using similar considerations we were always able to restrict the ambiguity
to a multiple of the quantity (\ref{amb2}). Let us also stress that the
$\widetilde{H}_{34}$ part of the 3PN Hamiltonian is written in such a form
that there is no need to use the rule (\ref{erd1}) for individual derivatives
appearing in $\widetilde{H}_{34}$.

A method of applying the rule (\ref{erd4}) in regularizations of integrals can
be found in Section 6 of \cite{J97}. We have yet found that this method
(without modifications) is not able to give zero for all full divergences one
meets in our calculations.


\begin{thebibliography}{99}

\bibitem{IW95} B.\ R.\ Iyer and C.\ M.\ Will, Phys.\ Rev.\ D {\bf 52}, 6882
(1995).

\bibitem{B97} L.\ Blanchet, Phys.\ Rev.\ D {\bf 55}, 714 (1997).

\bibitem{JS97} P.\ Jaranowski and G.\ Sch\"afer, Phys.\ Rev.\ D {\bf 55}, 4712 
(1997).

\bibitem{D87a} T.\ Damour, in {\em Gravitation in Astrophysics}, edited by
B.\ Carter and J.\ B.\ Hartle (Plenum Press, New York, 1987), p.\ 3.

\bibitem{GII97} A.\ Gopakumar, B.\ R.\ Iyer, and S.\ Iyer, preprint (1997),
gr-qc/9703075.

\bibitem{D87} T.\ Damour, in {\em Three Hundred Years of Gravitation}, edited
by S. Hawking and W. Israel (Cambridge University Press, Cambridge, 1987), p.\
128.

\bibitem{DS88} T.\ Damour and G.\ Sch\"afer, Nuovo Cimento B {\bf 101}, 127
(1988).

\bibitem{ADM62} R.\ Arnowitt, S.\ Deser, and C.\ W.\ Misner, in {\em
Gravitation: An Introduction to Current Research}, edited by L.\ Witten
(John Wiley, New York, 1962), p.\ 227.

\bibitem{KWW92/93} L.\ E.\ Kidder, C.\ M.\ Will, and A.\ G.\ Wiseman, Class.\
Quantum Grav. {\bf 9}, L125 (1992); Phys.\ Rev.\ D {\bf 47}, 3281 (1993).

\bibitem{WS93} N.\ Wex and G.\ Sch\"afer, Class.\ Quantum Grav.\ {\bf 10}, 2729
(1993).

\bibitem{SW93} G.\ Sch\"afer and N.\ Wex, in {\em Perspectives in Neutrinos,
Atomic Physics, and Gravitation}, edited by J.\ Tr{\^a}n Thanh V{\^a}n, T.\
Damour, E.\ Hinds, and J.\ Wilkerson (Editions Fronti{\`e}res, 1993), p.\ 513.

\bibitem{DIS97} T.\ Damour, B.\ R.\ Iyer, and B.\ S.\ Sathyaprakash, preprint
(1997), gr-qc/9705034.

\bibitem{D83} T.\ Damour, in {\em Gravitational Radiation}, edited by N.\
Deruelle and T.\ Piran (North-Holland, Amsterdam, 1983), p.\ 59.

\bibitem{DS85} T.\ Damour and G.\ Sch\"afer, Gen.\ Rel.\ Grav. {\bf 17}, 879
(1985).

\bibitem{K85} S.\ M.\ Kopeikin, Sov.\ Astron.\ {\bf 29}, 516 (1985).

\bibitem{W91} S.\ Wolfram, {\em Mathematica. A System for Doing Mathematics by
Computer}, 2nd ed. (Addison-Wesley, Redwood City, 1991). 

\bibitem{S85} G.\ Sch\"afer, Annals of Physics (NY) {\bf 161}, 81 (1985).

\bibitem{K61} T.\ Kimura, Prog.\ Theor.\ Phys.\ {\bf 26}, 157 (1961).

\bibitem{RT74} T.\ Regge and C.\ Teitelboim, Annals of Physics (NY) {\bf 88},
286 (1974).

\bibitem{S95} G.\ Sch\"afer, in {\em Symposia Gaussiana}, Proceedings of the
2nd Gauss Symposium, Conference A: Mathematical and Theoretical Physics,
Munich, edited by  M.\ Behara, R.\ Fritsch, and R.\ G.\ Lintz (Walter de
Gruyter, Berlin, 1995), p.\ 667.

\bibitem{DS91} T.\ Damour and G.\ Sch\"afer, J.\ Math.\ Phys. {\bf 32}, 127
(1991).

\bibitem{IZ80} C.\ Itzykson and J.\ B.\ Zuber, {\em Quantum Field Theory}
(McGraw-Hill, New York, 1980), p.\ 416.

\bibitem{KT72} T.\ Kimura and T.\ Toiya, Prog.\ Theor.\ Phys.\ {\bf 48}, 316
(1972).

\bibitem{OOKH74} T.\ Ohta, H.\ Okamura, T.\ Kimura, and K.\ Hiida, Prog.\
Theor.\ Phys.\ {\bf 51}, 1220 (1974); Nuovo Cimento B {\bf 27}, 103 (1975).

\bibitem{J97} P.\ Jaranowski, in {\em Mathematics of Gravitation. Part II.
Gravitational Wave Detection}, Banach Center Publications, Vol.\ 41, Part II
(Institute of Mathematics, Polish Academy of Sciences, Warszawa, 1997), p.\ 55.

\bibitem{R49} M.\ Riesz, Acta Math.\ {\bf 81}, 1 (1949).

\bibitem{GS64} I.\ M.\ Gel'fand and G.\ E.\ Shilov, {\it Generalized
Functions}, Vol.\ 1 (Academic Press, New York, 1964).

\end{thebibliography}
\end{document}